\documentstyle[aps,epsfig]{revtex}

\begin{document}

\title{When do tracer particles dominate the Lyapunov spectrum?}

\author{Pierre Gaspard}
\address{ Center for Nonlinear Phenomena and Complex Systems, Universit\'e
Libre de Bruxelles,
\\ Campus Plaine, Code Postal 231, 1050 Brussels, Belgium}

\author{Henk van Beijeren}

\address{Center for Nonlinear Phenomena and Complex Systems, Universit\'e
Libre de Bruxelles, \\
Campus Plaine, Code Postal 231, 1050 Brussels, Belgium \\ and \\ Institute
for Theoretical
Physics, Utrecht University, \\ Leuvenlaan 4, 3584 CE Utrecht, The Netherlands}

\date{\today}

\maketitle

\begin{abstract}
Dynamical instability is studied in a deterministic dynamical system of
Hamiltonian type
composed of a tracer particle in a fluid of many
particles.  The
tracer and fluid
particles are hard balls
(disks, in two dimensions,  or
spheres, in three dimensions) undergoing
elastic collisions.  The dynamical instability is characterized by the
spectrum of Lyapunov
exponents.  The tracer particle is shown to dominate the Lyapunov spectrum
in the neighborhoods of two limiting cases: the
Lorentz-gas limit in which the tracer particle is much lighter than the
fluid particles and the
Rayleigh-flight limit in which the fluid particles have a vanishing radius
and form an ideal
gas.  In both limits, a gap appears in the Lyapunov spectrum between the
few largest Lyapunov
exponents associated with the tracer and the rest of the Lyapunov spectrum.
\end{abstract}

\section{Introduction}

During the last decade,
dynamical instability and chaos in systems of
interacting
particles has become a problem of major preoccupation in statistical
mechanics.  Many systems
have been shown to present sensitivity to initial conditions characterized
by positive Lyapunov
exponents \cite{dorfman99,dellago96}.  Methods from kinetic theory have been
developed for the maximal Lyapunov exponent and the Kolmogorov-Sinai
entropy of dilute gases
\cite{vanbeijeren95,vanbeijeren97,vanbeijeren98,vanzon98,szasz,vanbeijeren99,van
beijeren01}.
The relationships
to transport properties have also been investigated
\cite{evans90,gaspard90,dorfman95,gaspard98}.

In mixtures of a one-component fluid with a very low concentration of
identical tracer particles the largest Lyapunov exponent may either be
virtually identical to the largest Lyapunov exponent of the pure fluid
(with just a slight perturbation caused by the tracer particles), or it
may be larger, due to the dynamical properties of the tracer particles. In
the latter case one may say that the largest Lyapunov exponent is {\em
dominated by the tracer particles}.

The purpose of the present paper is to describe the regimes in which the
tracer particles dominate the dynamical instability of the fluid system.
If the mass of the tracer particle is not much
larger than that of the fluid particles, and in addition the mean free
path of both bath and tracer particles are at least comparable in length
to their respective radii, the characteristic Lyapunov exponents for fluid
and tracer dynamics respectively are roughly of the order of their
respective collision frequencies.
So in this case one may conclude that the tracer particle will dominate if
its collision frequency is sufficiently larger than that of the fluid
particles.

One obvious way to reach this goal is by making the mass ratio
$M/m$ of tracer mass over fluid particle mass very small. The limit where
this ratio goes to zero corresponds to the Lorentz gas, for which the
Lyapunov exponents of the tracer particle were derived several years
ago~\cite{vanbeijeren95,vanbeijeren98,vanbeijeren99}.

Another common situation is where the radius of the tracer
particles is much larger than that of the fluid particles, e.g., when one
satisfies
the conditions for Brownian motion,
where also the mass of the tracer particle is much larger than that of the
fluid
particles. In this case the collision frequency
of the tracer particles is much higher than that of the fluid particles,
but at the same time the velocity changes of the tracer particle in
collisions with a bath particle are much smaller than those of the bath
particles themselves. In addition the curvature
of the Brownian particle, due to its large radius, is much smaller than
that of a fluid particle and therefore the diverging effect (crucial for a
positive Lyapunov exponent) of a Brownian-fluid collision is much smaller
than that of a fluid-fluid collision. As a result of this the largest
Lyapunov exponent usually is determined by the fluid particles, as noted
both by Louis and Gaspard~\cite{browngas} and by Nasser and
Dorfman~\cite{browndorf}. However,
there is one noteworthy exception to
this, corresponding to the so-called Rayleigh-flight
limit, where at fixed
Brownian mass and radius the radius of the fluid particles is sent to
zero. In the extreme limit of a Brownian gas surrounded by an ideal gas it
is obvious that the Brownian particle has to dominate, as the Lyapunov
exponents of the ideal gas are strictly zero.  The goal of the present
paper is thus to study
these cases of dominance of the Lyapunov spectrum by the tracer particle.

The plan of the paper is the following.  In Sec. \ref{dyninst}, the problem of
dynamical instability in a system of hard balls of different masses and
radii is
posed.  In Sec. \ref{lorentz}, we consider the Lorentz-gas limit.
In Sec. \ref{rayleigh}, we consider the Rayleigh-flight limit.
Conclusions are drawn in Sec. \ref{conclusions}.

\section{The dynamical system and its instability}
\label{dyninst}

\subsection{The dynamics}

In this paper we will consider a system composed of many fluid particles and
one tracer particle.  From a general viewpoint, this system can be
considered as
a binary mixture of $N_{\rm f}$ fluid particles, which
we take to be hard spheres of
radius
$a$ and mass $m$, with $N_{\rm t}$ tracer
particles, which
will likewise be hard spheres, with radius $A$ and mass $M$.  All
the $N=N_{\rm f}+N_{\rm t}$ hard balls move in a rectangular domain of finite
extension $\Omega$ with periodic boundary conditions.

The motion of the hard balls is composed of free flights between binary
collisions which are elastic and instantaneous.  Energy and the total
linear momenta are conserved.

For a system of hard balls of radii $\lbrace a_i\rbrace_{i=1}^{N}$ and of
masses
$\lbrace m_i\rbrace_{i=1}^{N}$, the equations of motion are given as follows in
terms of the positions and velocities $\lbrace {\bf r}^{(-)}_i(t_n),{\bf
v}^{(-)}_i(t_n)\rbrace_{i=1}^{N}$ and $\lbrace {\bf r}^{(+)}_i(t_n),{\bf
v}^{(+)}_i(t_n)\rbrace_{i=1}^{N}$, respectively before and after the
collision at time $t_n$:

\subsubsection{Free flight between binary collisions:}

\begin{equation}
t_{n-1}\to t_n: \quad \cases{{\bf r}_i^{(-)}(t_n) \ = \ {\bf
r}_i^{(+)}(t_{n-1})
\ + \ (t_n-t_{n-1}) \ {\bf v}_i^{(+)}(t_{n-1}) \cr
{\bf v}_i^{(-)}(t_n) \ = \ {\bf v}_i^{(+)}(t_{n-1}) \cr}
\label{free}
\end{equation}

\subsubsection{Binary collision:}

\begin{equation}
 t_n: \quad {\bf r}_i^{(+)} \ = \ {\bf r}_i^{(-)}
\label{r.coll}
\end{equation}
\begin{equation}
t_n: \quad \cases{{\bf v}_i^{(+)} \ = \ {\bf v}_i^{(-)}\ - \ 2 \
\frac{m_j}{m_i+m_j}\
(\hbox{\boldmath{$\epsilon$}}_{ij}\cdot {\bf v}_{ij}^{(-)}) \
\hbox{\boldmath{$\epsilon$}}_{ij}
\cr {\bf v}_j^{(+)} \ = \ {\bf v}_j^{(-)}\ + \ 2 \ \frac{m_i}{m_i+m_j}\
(\hbox{\boldmath{$\epsilon$}}_{ij}\cdot {\bf v}_{ij}^{(-)}) \
\hbox{\boldmath{$\epsilon$}}_{ij}
\cr{\bf v}_k^{(+)} \ = \ {\bf v}_k^{(-)} \qquad\qquad{\rm for} \quad k\not=
i,j \cr}
\label{v.coll}
\end{equation}
with the unit vector joining the centers of the $i^{\rm th}$ and $j^{\rm
th}$ balls at the collision given by
\begin{equation}
\hbox{\boldmath{$\epsilon$}}_{ij} \ \equiv\ \frac{{\bf r}_i^{(\pm)}-{\bf
r}_j^{(\pm)}}{a_i+a_j}
\label{e.coll}
\end{equation}
and the relative velocity vector
\begin{equation}
{\bf v}_{ij}^{(-)} \ \equiv\ {\bf v}_i^{(-)}-{\bf v}_j^{(-)}
\label{vv.coll}
\end{equation}

\subsection{The linearized dynamics}

The dynamical instability of this system is characterized by the rates of
exponential growth of infinitesimal perturbations on the positions and
velocities of the particles: $\delta{\bf X}=\lbrace \delta{\bf r}_i,\delta{\bf
v}_i\rbrace_{i=1}^{N}$.  The rates are called the Lyapunov exponents
\begin{equation}
\lambda = \lim_{t\to\infty} \frac{1}{t}\ln\frac{\Vert\delta{\bf
X}(t)\Vert}{\Vert\delta{\bf X}(0)\Vert}
\end{equation}
Depending on the initial perturbation $\delta{\bf X}(0)$ we may have as many
different Lyapunov exponents as there are
independent directions in phase space.  Since the
dynamics of the present hard-ball system is of Hamiltonian (symplectic)
character the Lyapunov exponents obey a pairing rule:  {\it If $\lambda_i$ is a
Lyapunov exponent, then $-\lambda_i$ is also a Lyapunov exponent}.  The set
of exponents form
the so-called Lyapunov spectrum.  The Lyapunov exponents associated with
the directions
perpendicular to the energy and momenta shells vanish.
In the present system, which has no fixed points except at zero energy,
this also holds for
the pair mates associated with the
directions of time and center-of-mass
translation.  Accordingly, $2+2d$
Lyapunov exponents
vanish.

The problem of the dynamical instability of a hard-ball system was
formulated in the seventies
by Sinai \cite{sinai70} who was inspired by the pioneering work of Krylov
in the forties
\cite{krylov}.  Independently,
Erpenbeck and Wood carried out numerical investigations,
also in the seventies \cite{erpenbeck}.  The systematic calculation of the
Lyapunov spectrum in
hard-ball systems has been developed in the
nineties.  Using the method of
Gaspard and Dorfman
\cite{gaspard95}
one can derive the following linearized equations
in terms
of the infinitesimal
perturbations
before and after each collision \cite{browngas}:

\subsubsection{Free flight between binary collisions:}

\begin{equation}
t_{n-1}\to t_n: \quad \cases{\delta{\bf r}_i^{(-)}(t_n) \ = \ \delta{\bf
r}_i^{(+)}(t_{n-1}) \ + \
(t_n-t_{n-1}) \ \delta{\bf v}_i^{(+)}(t_{n-1}) \cr
\delta{\bf v}_i^{(-)}(t_n) \ = \ \delta{\bf v}_i^{(+)}(t_{n-1}) \cr}
\label{d.free}
\end{equation}

\subsubsection{Binary collision:}

\begin{equation}
 t_n: \quad \cases{\delta{\bf r}_i^{(+)} \ = \ \delta{\bf r}_i^{(-)} \ - \
2 \ \frac{m_j}{m_i+m_j}
(\hbox{\boldmath{$\epsilon$}}_{ij}\cdot \delta{\bf r}_{ij}^{(-)}) \
\hbox{\boldmath{$\epsilon$}}_{ij}\cr
\delta{\bf r}_j^{(+)} \ = \ \delta{\bf r}_j^{(-)} \ + \ 2 \
\frac{m_i}{m_i+m_j}
(\hbox{\boldmath{$\epsilon$}}_{ij}\cdot \delta{\bf r}_{ij}^{(-)}) \
\hbox{\boldmath{$\epsilon$}}_{ij}\cr
\delta{\bf r}_k^{(+)} \ = \ \delta{\bf r}_k^{(-)} \qquad\qquad{\rm for}
\quad k\not= i,j
\cr}
\label{dr.coll}
\end{equation}
\begin{equation}
t_n: \quad \cases{\delta{\bf v}_i^{(+)} \ = \ \delta{\bf v}_i^{(-)}\ - \ 2
\ \frac{m_j}{m_i+m_j}\
\Bigl\lbrack (\hbox{\boldmath{$\epsilon$}}_{ij}\cdot \delta{\bf
v}_{ij}^{(-)}) \
\hbox{\boldmath{$\epsilon$}}_{ij} +
(\delta\hbox{\boldmath{$\epsilon$}}_{ij}\cdot {\bf
v}_{ij}^{(-)})\ \hbox{\boldmath{$\epsilon$}}_{ij} +
(\hbox{\boldmath{$\epsilon$}}_{ij}\cdot {\bf
v}_{ij}^{(-)})\ \delta\hbox{\boldmath{$\epsilon$}}_{ij}\Bigr\rbrack\cr
\delta{\bf v}_j^{(+)} \ = \ \delta{\bf v}_j^{(-)}\ + \ 2 \
\frac{m_i}{m_i+m_j}\
\Bigl\lbrack (\hbox{\boldmath{$\epsilon$}}_{ij}\cdot \delta{\bf
v}_{ij}^{(-)}) \
\hbox{\boldmath{$\epsilon$}}_{ij} +
(\delta\hbox{\boldmath{$\epsilon$}}_{ij}\cdot {\bf
v}_{ij}^{(-)})\ \hbox{\boldmath{$\epsilon$}}_{ij} +
(\hbox{\boldmath{$\epsilon$}}_{ij}\cdot {\bf
v}_{ij}^{(-)})\ \delta\hbox{\boldmath{$\epsilon$}}_{ij}\Bigr\rbrack\cr
\delta{\bf v}_k^{(+)} \ = \ \delta{\bf v}_k^{(-)} \qquad\qquad{\rm for}
\quad k\not= i,j  \cr}
\label{dv.coll}
\end{equation}
with
\begin{equation}
\delta\hbox{\boldmath{$\epsilon$}}_{ij} = \frac{1}{a_i+a_j} \Biggl(
\delta{\bf r}_{ij}^{(-)} \ - \
{\bf v}_{ij}^{(-)} \ \frac{\hbox{\boldmath{$\epsilon$}}_{ij}\cdot \delta{\bf
r}_{ij}^{(-)}}{\hbox{\boldmath{$\epsilon$}}_{ij}\cdot{\bf v}_{ij}^{(-)}}\Biggr)
\label{de.coll}
\end{equation}
and
\begin{eqnarray}
\delta{\bf r}_{ij}^{(-)} \ &\equiv& \ \delta{\bf r}_i^{(-)}-\delta{\bf
r}_j^{(-)} \label{ddr.coll}\\
\delta{\bf v}_{ij}^{(-)} \ &\equiv& \ \delta{\bf v}_i^{(-)}-\delta{\bf
v}_j^{(-)}
\label{ddv.coll}
\end{eqnarray}

\subsection{The kinetic properties of the thermodynamic equilibrium state}

We require that the center of mass is at rest and we are
interested in the properties of the equilibrium thermodynamic state at fixed
temperature $T$.  Accordingly, the total linear momenta vanish:
${\bf P}_{\rm
tot}=\sum_{i=1}^N m_i{\bf v}_i=0$,
while the total energy is given by
\begin{equation}
E=\sum_{i=1}^N \frac{1}{2}m_i{\bf v}_i^2 = \frac{d}{2}(N-1)k_{\rm B}T
\end{equation}
where $k_{\rm B}$ is Boltzmann's constant.  For hard-ball systems, the
motions at different temperatures are equivalent up to a rescaling of time.  In
the sequel, the temperature is thus fixed at the value $T=k_{\rm B}^{-1}$.

In the fluid phase at low enough density, the system is supposed to be
ergodic on
each energy-momenta shell, which defines the equilibrium states.
The mean
velocity\footnote{
The mean velocity should be well distinguished from the root-mean-square
velocity,
defined as $\sqrt{\langle {\bf v}_i^2\rangle}$.} of each particle is defined as
\begin{equation}
v_i\equiv \langle \Vert{\bf v}_i\Vert\rangle
\end{equation}
At equilibrium, the mean velocities are determined by the temperature and the
mass of the particles.
For large particle number, they are given, to leading order in $1/N$, by
\begin{eqnarray}
&d=2:& \qquad  v_i= \sqrt{\frac{\pi k_{\rm B}T}{2 m_i}} \label{v2}\\
&d=3:& \qquad  v_i= \sqrt{\frac{8 k_{\rm B}T}{\pi m_i}} \label{v3}
\end{eqnarray}
where $d$ is the space dimension.

The mean relative velocities between the particles entering a binary collision
will also be of importance
in the sequel. They are defined by
\begin{equation}
v_{ij}\equiv \langle \Vert{\bf v}_i-{\bf v}_j\Vert\rangle
\end{equation}
and are given by Eqs. (\ref{v2})-(\ref{v3}) with the mass replaced by the
relative mass
\begin{equation}
\mu_{ij}\equiv \frac{m_im_j}{m_i+m_j}
\end{equation}

If the system is sufficiently dilute the collision frequencies of the fluid and
tracer particles can be evaluated by supposing that each particle has a
cross-section for collision with each one of the other types of particles:
\begin{eqnarray}
&d=2:& \qquad  \sigma_{ij}= 2 \, (a_i+a_j) \label{cs2}\\
&d=3:& \qquad  \sigma_{ij}= \pi \, (a_i+a_j)^2 \label{cs3}
\end{eqnarray}

If $\sigma_{\rm ff}$, $\sigma_{\rm ft}=\sigma_{\rm tf}$, and $\sigma_{\rm tt}$
denote the cross-sections for fluid-fluid, fluid-tracer, tracer-fluid, and
tracer-tracer collisions, the collision
frequencies of the tracer and fluid
particles
to leading order in the fluid density are
given by
\begin{eqnarray}
\nu_{\rm t}&=& \nu_{\rm tf}+\nu_{\rm tt} \simeq \frac{N_{\rm f}}{\Omega}\;
\sigma_{\rm tf} \; v_{\rm tf} + \frac{N_{\rm t}-1}{\Omega}
\; \sigma_{\rm tt} \; v_{\rm tt} \label{nu_t}\\
\nu_{\rm f}&=& \nu_{\rm ff}+\nu_{\rm ft} \simeq \frac{N_{\rm f}}{\Omega}
\; \sigma_{\rm ff} \; v_{\rm ff} + \frac{N_{\rm t}}{\Omega}
\; \sigma_{\rm ft} \; v_{\rm ft}\label{nu_f}
\end{eqnarray}
where $v_{\rm tf}=v_{\rm ft}$ and the extension parameter
$\Omega$ is the area respectively the volume of the system:
\begin{eqnarray}
&d=2:& \qquad  \Omega= L_x \; L_y \label{O2}\\
&d=3:& \qquad  \Omega= L_x \; L_y \; L_z \label{O3}
\end{eqnarray}
We notice that each tracer particle may collide
on the $N_{\rm t}-1$ other tracer particles, which explains
the presence of $-1$ in the tracer-tracer collision frequency $\nu_{\rm tt}$.
The term $-1$ is important in systems with a low number $N_{\rm t}$
of tracer particles.  In contrast, it can be neglected if the number
of particles is large as it is the case for the fluid particles.

\subsection{Simulations and system preparation}
In order to test our theoretical results we performed several MD
simulations in
which we computed the Lyapunov spectra of hard ball systems containing a
tracer
component.
In all these simulations, we consider a fluid with a single tracer particle,
so $N_{\rm t}=1$.

To initialize a simulation
we locate the fluid particles
on the lattice points of a crystal lattice.  In $d=2$, the initial positions
form a triangular lattice with $M_xM_y$ rectangular cells of two disks each.
In $d=3$, the initial positions form a face-centered cubic (FCC) lattice with
$M_xM_yM_z$ cubic cells of four spheres each.  The sizes
of the cells are fixed in order for the fluid to have a fixed density $n$
in absence of the tracer particle.  In $d=2$, the domain is rectangular of
sizes
\begin{equation}
d=2: \qquad L_x=M_x\; \left( \frac{2}{n\sqrt{3}}\right)^{\frac{1}{2}} \quad
\mbox{and} \quad
L_y=M_y\; \left( \frac{2\sqrt{3}}{n}\right)^{\frac{1}{2}}
\end{equation}
so that its area is $\Omega=L_xL_y=2M_xM_y/n$.  In $d=3$, the domain is
also rectangular but of
sizes
\begin{equation}
d=3: \qquad L_x=M_x\; \left( \frac{4}{n}\right)^{\frac{1}{3}} \ , \quad
L_y=M_y\; \left(
\frac{4}{n}\right)^{\frac{1}{3}} \ , \quad \mbox{and}
\quad L_z=M_z\; \left( \frac{4}{n}\right)^{\frac{1}{3}}
\end{equation}
so that its volume is $\Omega=L_xL_yL_z=4M_xM_yM_z/n$.

Thereafter, the tracer particle is placed in
the middle of the
crystal configuration
under removal of all the fluid particles that
would overlap with the tracer particle.  The number of fluid particles
is thus given approximately by
\begin{equation}
N_{\rm f} \simeq n \; (\Omega-\Omega_{\rm t})
\end{equation}
where
\begin{eqnarray}
&d=2:& \qquad  \Omega_{\rm t}= \pi\; A^2 \label{Ot2}\\
&d=3:& \qquad  \Omega_{\rm t}= \frac{4\pi}{3} \; A^3 \label{Ot3}
\end{eqnarray}
As a result of this procedure the density of the fluid particles in the
{\it free
volume} is still given to an excellent approximation by $n$, even if the
tracer
particle is so large that it occupies an appreciable fraction of the total
volume $\Omega$.

Accordingly, the collision frequency of the tracer particle is
\begin{equation}
\nu_{\rm t}= n \; \sigma_{\rm tf} \; v_{\rm tf}
\label{collfreq_t}
\end{equation}
and its mean free path is
\begin{equation}
\ell_{\rm t}= \frac{v_{\rm t}}{\nu_{\rm t}} \simeq
\frac{v_{\rm t}}{n \; \sigma_{\rm tf} \; v_{\rm tf}}
\end{equation}
Table I gives these quantities in $d=2$ and $d=3$.

\vskip 1 cm
\hrule \vskip1pt \hrule height1pt
\vskip 0.1 cm

\begin{center}
{ Table I. Characteristic quantities for the motion of a tracer in a dilute
fluid.}
\end{center}

\vskip 0.1 cm
\hrule
\vskip 0.1 cm

$$
\vcenter{\openup1\jot \halign{#\hfil&\qquad#\hfil&\qquad#\hfil\cr
{\it dimension} & $d=2$ & $d=3$ \cr
{\it mean velocity} & $v_{\rm t}=\sqrt{\frac{\pi k_{\rm B}T}{2M}}$ &
$v_{\rm t}=\sqrt{\frac{8 k_{\rm B}T}{\pi M}}$\cr
{\it collision frequency} &
$\nu_{\rm t} \simeq 2\, (a+A) \, n \, \sqrt{\frac{\pi k_{\rm B}T(m+M)}{2mM}}$ &
$\nu_{\rm t} \simeq \pi\, (a+A)^2 \, n \, \sqrt{\frac{8 k_{\rm
B}T(m+M)}{\pi mM}}$ \cr
{\it mean free path} &
$\ell_{\rm t} =\frac{v_{\rm t}}{\nu_{\rm t}}\simeq \frac{1}{2\, (a+A) \, n}
\sqrt{\frac{m}{m+M}}$&
$\ell_{\rm t} =\frac{v_{\rm t}}{\nu_{\rm t}} \simeq \frac{1}{\pi\, (a+A)^2
\, n}
\sqrt{\frac{m}{m+M}}$ \cr }}$$

\vskip 0.1 cm
\hrule height1pt \vskip1pt \hrule
\vskip 1 cm
Finally, the initial velocities of the particles are drawn with a random
number
generator from a Maxwellian distribution with the appropriate mass. In
practice
we choose $k_BT=1$ as well as $m=1$.

\subsection{Discussion of the dominance of the dynamical instability by the
tracer}

Since there is a single tracer particle it only collides with the fluid
particles.
However, each fluid particle may collide both with other fluid particles
and with the tracer particle.  Accordingly, there are two types of
collisions: the
fluid-fluid
collisions and the tracer-fluid collisions.

Since both types of collisions happen between particles with convex
surfaces, we
expect that both of them will contribute to the dynamical instability. As
discussed in the introduction, the strongest instability, characterized by the
largest Lyapunov exponent, could be dominated either by the fluid-fluid
collisions, or by the tracer-fluid collisions. To decide which of these
possibilities is realized for a given choice of parameters we can proceed
in the
following way: We calculate
both the fluid-fluid and the tracer-fluid maximal Lyapunov exponent {\em under
the assumption that indeed fluid-fluid collisions repectively tracer-fluid
collisions are dominant} and then compare the results. In the great
majority of
cases indeed the larger of the two calculated exponents gives an excellent
approximation to the actual
value of the maximal Lyapunov exponent.

The first calculation concerns the fluid-fluid collisions.  In the absence of
tracer, such collisions give the maximal Lyapunov exponent
\begin{equation}
\lambda_{\rm f} \simeq \omega(N_{\rm f}) \; \nu_{\rm f} \; \ln
\left[\frac{\alpha(N_{\rm
f})}{4\, n\, a^d}\right] \qquad \mbox{with}\quad \nu_{\rm f}\simeq n\;
\sigma_{\rm ff}\; v_{\rm
ff}
\label{fluidlyap}
\end{equation}
where, for disks in $d=2$, Van Zon {\it et al.} \cite{vanzon98} have shown that
the prefactor $\omega(N)$ is well fitted by the expression
\begin{equation}
\omega(N) \simeq 4.311 - \frac{3.466}{N^{0.277}},
\end{equation}
although the actual asymptotic behavior for large $N$ is as $1/(\log
N)^2$~\cite{szasz}.
Further, $\alpha(N_{\rm f})$ is of order unity, depends only weakly on
$N_{\rm f}$
and approaches a constant for $N_{\rm f} \to \infty$.

On the other hand, we may
consider the motion of the tracer particle undergoing kicks
by independent fluid
particles. The equations of motion are given by Eqs.
(\ref{free})-(\ref{vv.coll})
in which ${\bf r}_{i=1}={\bf R}$, ${\bf v}_{i=1}={\bf V}$.  We denote by $({\bf
R}_n^{(-)},{\bf V}_n^{(-)})$ and $({\bf R}_n^{(+)},{\bf V}_n^{(+)})$ the
position and
velocity of the tracer particle before
repectively after the $n^{\rm th}$ collision,
which occurs
at the time $t_n$. Since the trajectory is continuous in position,
${\bf
R}_n^{(+)}={\bf
R}_n^{(-)}\equiv {\bf R}_n$. On the other hand,
${\bf V}_n^{(+)}={\bf V}_{n+1}^{(-)}\equiv {\bf V}_n$.  We have the following
iteration for the motion itself:
\begin{eqnarray}
{\bf R}_n&=&{\bf R}_{n-1}+\tau_n\, {\bf V}_{n-1} \\
{\bf V}_n&=&{\bf V}_{n-1}- \frac{2\, m}{m+M} \, \left[
\hbox{\boldmath{$\epsilon$}}_n\cdot({\bf
V}_{n-1}-{\bf v}_{n-1})\right] \, \hbox{\boldmath{$\epsilon$}}_n \\
t_n&=&t_{n-1}+\tau_n
\end{eqnarray}
where $\tau_n$ is the time interval between the $(n-1)^{\rm th}$ and
the
$n^{\rm th}$ collision,
while
\begin{equation}
\hbox{\boldmath{$\epsilon$}}_n \equiv \frac{{\bf R}_n-{\bf r}_n}{A+a}
\end{equation}
is the unit impact vector at the $n^{\rm th}$ collision.

An infinitesimal perturbation on the motion is ruled by Eqs.
(\ref{d.free})-(\ref{ddv.coll}) in which $\delta{\bf r}_{i=1}=\delta{\bf
R}$, $\delta{\bf
v}_{i=1}=\delta{\bf V}$ and $\delta{\bf r}_j^{(\pm)}=\delta{\bf
v}_j^{(\pm)}= 0$ for
$j\neq 1$ because,
if indeed the tracer particle dominates the maximal Lyapunov exponent, the
perturbations of the fluid particle positions and velocities will be negligibly
small compared to those of the tracer particle. For this argument to hold it is
important that recollisions of a fluid particle with the tracer particle are
either rare, or, if they are not, are still dominated by the perturbations
of the
tracer particle. Especially when $M\gg m$ this is a subtle point, because,
as can
be seen from (\ref{d.free})-(\ref{ddv.coll}), the perturbations of a fluid
particle
right after a collision are comparable to those of the tracer particle. So if a
recollision is not unlikely, it has to occur typically after a time that is
long,
compared to the Lyapunov time of the tracer particle. The iteration for an
infinitesimal perturbation on the motion is
then given by
\begin{eqnarray}
\mbox{\it free flight:} \qquad
\delta{\bf R}_n^{(-)}&=&\delta{\bf R}_{n-1}^{(+)}+\tau_n\; \delta{\bf
V}_{n-1}
\label{dR}\\
\mbox{\it binary collision:} \qquad
\delta{\bf R}_n^{(+)}&=&\delta{\bf R}_n^{(-)}- \frac{2\, m}{m+M}
\left(\hbox{\boldmath{$\epsilon$}}_n\cdot\delta{\bf R}_n^{(-)}\right)
\; \hbox{\boldmath{$\epsilon$}}_n
\label{dR2}\\
\delta{\bf V}_n&=&\delta{\bf V}_{n-1}-\frac{2\, m}{m+M} \, \left\{
(\hbox{\boldmath{$\epsilon$}}_n\cdot\delta{\bf V}_{n-1}) \,
\hbox{\boldmath{$\epsilon$}}_n +
\left[\delta\hbox{\boldmath{$\epsilon$}}_n\cdot({\bf
V}_{n-1}-{\bf v}_{n-1})\right]\, \hbox{\boldmath{$\epsilon$}}_n \right.
\nonumber\\
 & &  \left. \qquad \qquad \qquad \qquad \qquad \qquad \qquad \quad +
\left[ \hbox{\boldmath{$\epsilon$}}_n\cdot({\bf
V}_{n-1}-{\bf v}_{n-1})\right] \, \delta\hbox{\boldmath{$\epsilon$}}_n\right\}
\label{dV}
\end{eqnarray}
with $\delta{\bf V}_n=\delta{\bf V}_n^{(+)}=\delta{\bf V}_{n+1}^{(-)}$ and
\begin{equation}
\delta\hbox{\boldmath{$\epsilon$}}_n=\frac{1}{A+a} \left[
\delta{\bf R}_n^{(-)}- ({\bf V}_{n-1}-{\bf v}_{n-1}) \;
\frac{\hbox{\boldmath{$\epsilon$}}_n\cdot\delta{\bf R}_n^{(-)}}
{\hbox{\boldmath{$\epsilon$}}_n\cdot({\bf
V}_{n-1}-{\bf v}_{n-1})} \right]
\label{de}
\end{equation}
In the equations
above, the perturbation
$\delta{\bf R}_n^{(-)}$
corresponds to the
first impact time of the two neighboring trajectories and the perturbation
$\delta{\bf R}_n^{(+)}$ to the later impact time.
They are related as
\begin{equation}
\delta{\bf R}_n^{(+)} = \delta{\bf R}_n^{(-)}+
({\bf V}_{n-1}-{\bf V}_{n})\, \delta\tau_n \, ;
\end{equation}
with the time lag at collision
\begin{equation}
\delta\tau_n= - \frac{\hbox{\boldmath{$\epsilon$}}_n\cdot\delta{\bf R}_n^{(-)}}
{\hbox{\boldmath{$\epsilon$}}_n\cdot({\bf
V}_{n-1}-{\bf v}_{n-1})}
\label{dt}
\end{equation}

These equations can be analyzed thanks to a few observations.  First of
all, the time $\tau_n$
between two successive collisions has an average value given
as the inverse of
the collision
frequency of the tracer particle:
\begin{equation}
\langle \tau_n\rangle = \frac{1}{\nu_{\rm t}}
\end{equation}
Furthermore, we find the mean relative velocity as
\begin{equation}
\langle \Vert{\bf V}_{n-1}-{\bf v}_{n-1}\Vert\rangle = v_{\rm tf}
\end{equation}

We may expect a dominance
of the resulting Lyapunov exponents for the tracer particle over
those resulting from the
fluid-fluid collisions in two cases to be discussed below:
in the Lorentz-gas limit $M\to 0$ with $M\ll m$, and in the
Rayleigh-flight limit $a\to 0$ with $a\ll A$ and $na^d\ll 1$.
The calculation of these Lyapunov exponents will be the subject of the next two
sections.

\section{The Lorentz-gas limit}
\label{lorentz}

In the limit $M\to 0$,
with $M \ll m$, the tracer particle is much faster
than the fluid
particles.  Accordingly, the tracer moves
through a fluid which is essentially at
rest.  This system is referred to as a Lorentz gas
\cite{lorentz05,vanbeijeren82}.  The
collision frequency of the tracer particle will thus be larger than the
collision frequency of
the fluid particles.  Therefore, the perturbations on the coordinates of
the tracer particle
will grow faster than the perturbations on the fluid particles.  Actually,
the value of the
maximal Lyapunov exponent of such a fluid can be predicted in this case
thanks to the
work by Van Beijeren,
Dorfman, and Latz \cite{vanbeijeren95,vanbeijeren98}.

\subsection{The two-dimensional case}

In $d=2$, the Lorentz gas has a single positive Lyapunov exponent as a
consequence of
chaos and energy conservation. Therefore, we expect that the maximal
Lyapunov exponent is the
positive Lyapunov exponent of the Lorentz gas when $M\ll m$,
while the next Lyapunov exponents
remain at the level of the fluid Lyapunov exponent (\ref{fluidlyap}).  For
a dilute system with
$M\ll m$, the maximal Lyapunov exponent is therefore given by
\begin{equation}
d=2:\qquad \lambda_1 \simeq 2\, (a+A)\, n \, \sqrt{\frac{\pi k_{\rm
B}T(m+M)}{2mM}}
\; \ln\left[ \frac{{\rm e}^{1-{\cal C}}}{2n(a+A)^2}\right]
\label{Lor2}
\end{equation}
with Euler's constant ${\cal C}=0.5772156649...$ \cite{vanbeijeren95}.

\begin{figure}
\centerline{\epsfig{file=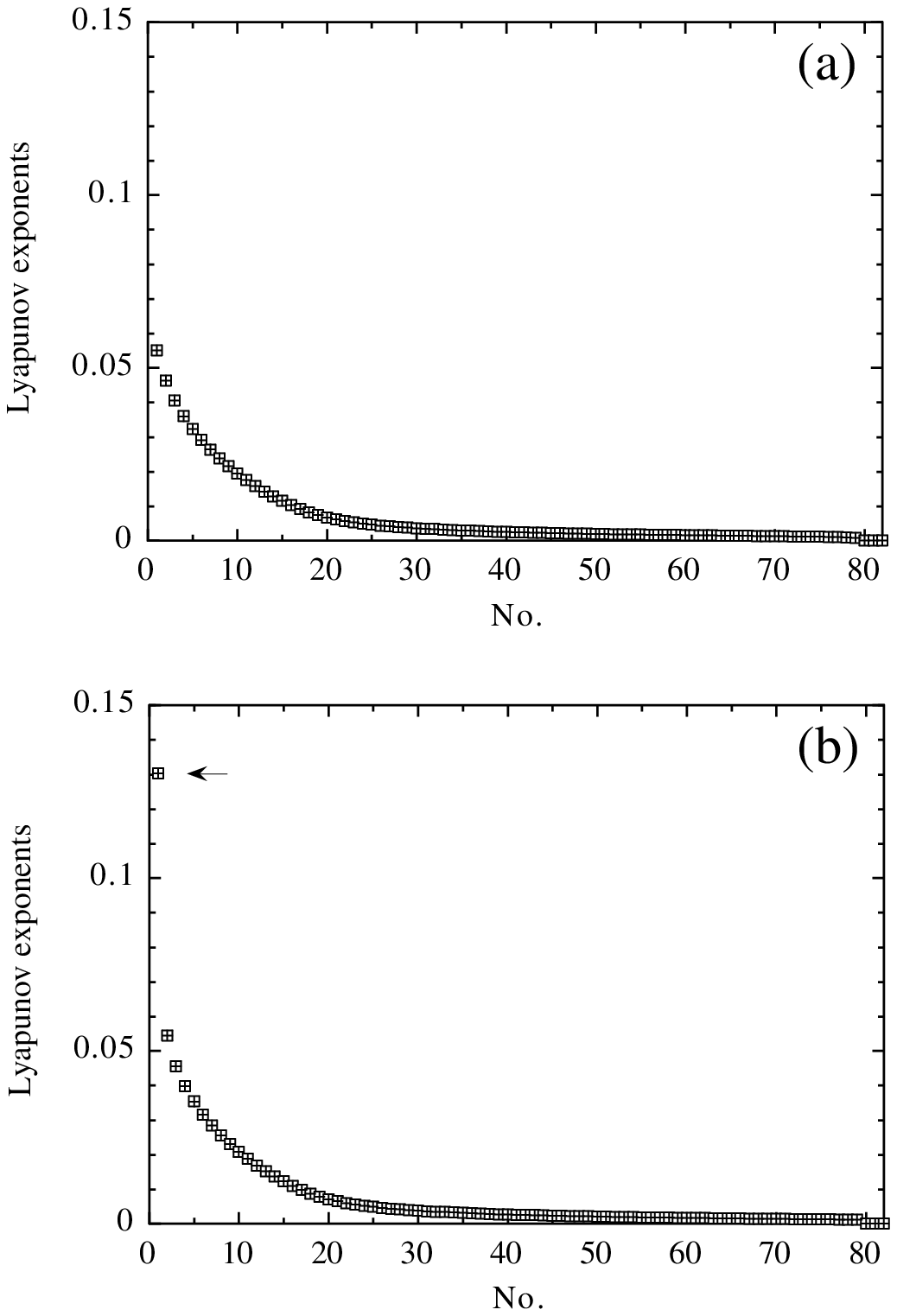,width=10cm}}
\caption{Lyapunov spectra of a 2d fluid of density $n=10^{-3}$ and temperature
$T=1$ composed of $N_{\rm f}=40$ hard disks of radius $a=1/2$ and mass
$m=1$ with $N_{\rm t}=1$
hard disk of radius $A=1/6$ and mass: (a) $M=10$; (b) $M=10^{-2}$.  The squares
depict the positive exponents and the crosses are minus the negative
exponents.}
\label{fig1}
\end{figure}

This behavior is
well confirmed by numerical computation.
Figure \ref{fig1}
compares Lyapunov spectra for $M\gg m$ and for $M\ll m$ at a fixed ratio of
$a$ and $A$.
We observe that the maximal Lyapunov exponent is separated from the rest of
the spectrum for $M\ll m$.
This implies that with increasing $m/M$ a gap opens up
in the Lyapunov spectrum.

\begin{figure}
\centerline{\epsfig{file=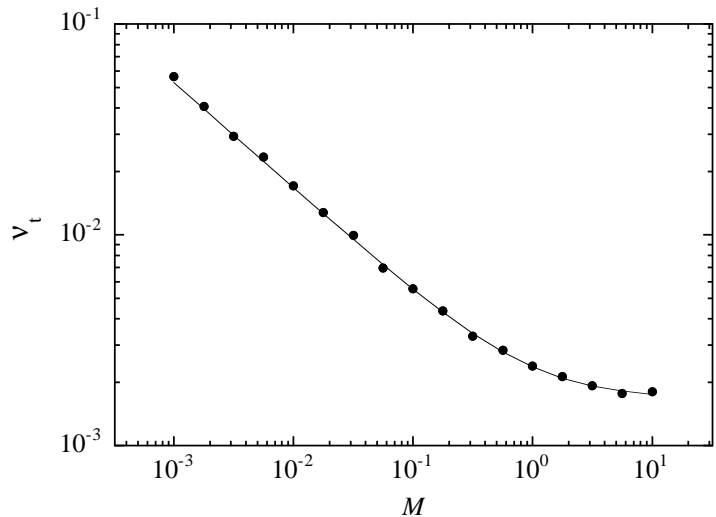,width=10cm}}
\caption{Collision frequency $\nu_{\rm t}$ of the tracer particle in a 2d
fluid of density
$n=10^{-3}$ and temperature $T=1$ composed of $N_{\rm f}=40$ hard disks of
radius $a=1/2$ and
mass $m=1$ with $N_{\rm t}=1$ hard disk of radius $A=1/6$ and varying mass
$M$.  The filled
circles are the numerical data.  The solid line is the prediction of Eq.
(\ref{collfreq_t})
(see Table I).}
\label{fig2}
\end{figure}

Figure \ref{fig2} shows that
in the limit $M\to 0$ the collision frequency of the tracer particle
increases as
predicted by Eq. (\ref{collfreq_t}) in the limit $M\to 0$.

\begin{figure}
\centerline{\epsfig{file=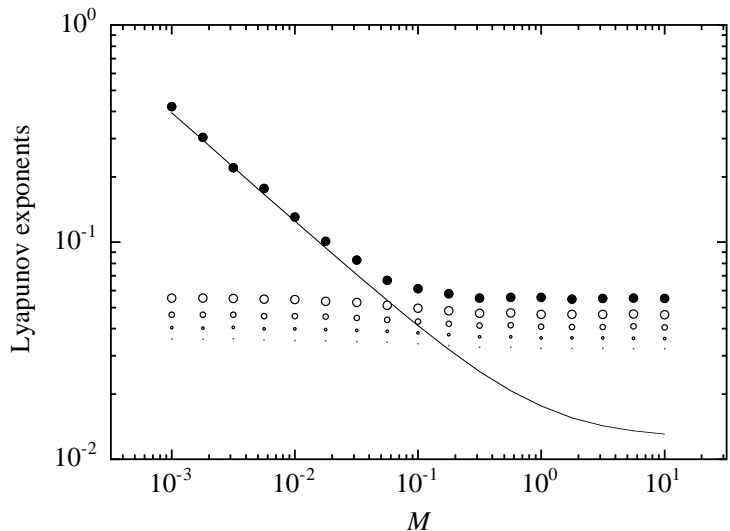,width=10cm}}
\caption{The five largest Lyapunov exponents of a 2d fluid of density
$n=10^{-3}$ and
temperature $T=1$ composed of $N_{\rm f}=40$ hard disks of radius $a=1/2$
and mass $m=1$ with
$N_{\rm t}=1$ hard disk of radius $A=1/6$ and varying mass $M$.  The filled
circles depict the
largest Lyapunov exponent and the open circles the four next ones.  The
solid line is the
prediction of Eq. (\ref{Lor2}).}
\label{fig3}
\end{figure}

Figure \ref{fig3} depicts the five largest Lyapunov exponents as
a function of the mass of the tracer particle showing that, indeed, the
Lyapunov spectrum is
dominated by the tracer particle as soon as $M\ll m$.  The maximal Lyapunov
exponent undergoes a
cross-over from the fluid value (\ref{fluidlyap})
for $M \gtrsim m$
to the
Lorentz-gas value
(\ref{Lor2}) for $M\ll m$.  In the regime $M\ll m$, the second Lyapunov
exponent
tends to
slightly increase up to the fluid value (\ref{fluidlyap}) $\lambda_2\simeq
\lambda_{\rm f}<\lambda_1$.  A similar behavior is seen for the next
Lyapunov exponents.

\begin{figure}
\centerline{\epsfig{file=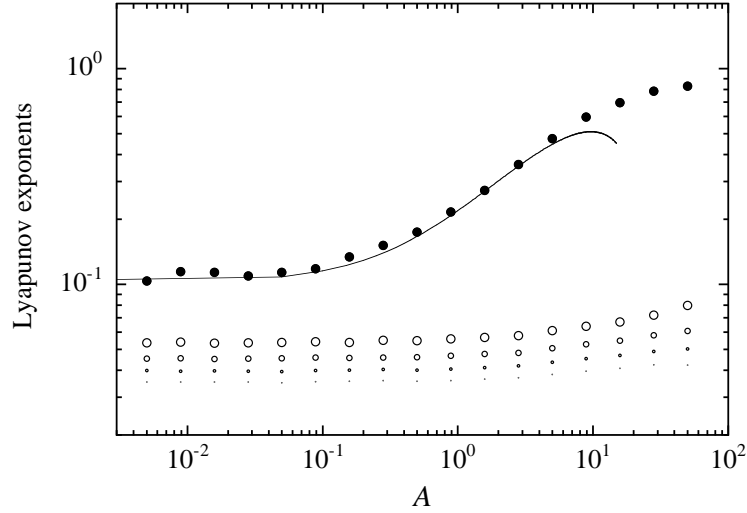,width=10cm}}
\caption{The five largest Lyapunov exponents of a 2d fluid of density
$n=10^{-3}$ and
temperature $T=1$ composed of $N_{\rm f}$ hard disks of radius $a=1/2$ and
mass $m=1$ with
$N_{\rm t}=1$ hard disk of varying radius $A$ and mass $M=10^{-2}$ ($N_{\rm
f}=40$ for $A<10$,
$N_{\rm f}=39$, $37$, and $31$ for the last three values of $A>10$).  The
filled circles depict
the largest Lyapunov exponent and the open circles the four next ones.  The
solid line is the
prediction of Eq. (\ref{Lor2}).}
\label{fig4}
\end{figure}

Figure \ref{fig4} shows the dependence of the spectrum on the radius $A$ of
the tracer
particle.  As the tracer becomes larger and larger in a fluid of fixed
density, the room for its
motion between the fluid particles
becomes relatively smaller and smaller.
A cage effect occurs for a large tracer as though the
effective density of the Lorentz gas would increase
(indeed the motion of a mobile particle of radius $A$ among fixed point
scatterers is completely equivalent to that of a
mobile point particle among fixed scatterers of radius $A$ that are
allowed to overlap each other. In the literature this
has been treated as the Lorentz gas with overlapping scatterers).
In this case, the assumption $\ell_{\rm t}/A \gg 1$, on
which the demonstration of (\ref{Lor2}) is based, breaks down,
which explains the discrepancy
observed in Fig. \ref{fig4} for
$A\gg a$ between the
numerical results and the prediction of Eq. (\ref{Lor2}).

\subsection{The three-dimensional case}

In $d=3$, the Lorentz gas has two positive Lyapunov exponents as a
consequence of chaos and
energy conservation. In this case, we thus expect that the two largest
Lyapunov exponents
are the
positive Lyapunov exponents of the Lorentz gas when $M\ll m$, while the next
Lyapunov exponents
remain at the levels
of the fluid Lyapunov exponents (\ref{fluidlyap}). For a
dilute system with
$M\ll m$, the two largest Lyapunov exponents are therefore given by
\cite{vanbeijeren98}
\begin{eqnarray}
d=3: \nonumber\\
\lambda_1 &\simeq&  \pi \, (a+A)^2 \, n \, \sqrt{\frac{8 k_{\rm
B}T(m+M)}{\pi mM}}\; \ln\left[
\frac{4\, {\rm e}^{-\frac{1}{2}-{\cal C}}}{n\pi(a+A)^3}\right] \label{1Lor3}\\
\lambda_2 &\simeq&  \pi \, (a+A)^2 \, n \, \sqrt{\frac{8 k_{\rm
B}T(m+M)}{\pi mM}}\; \ln\left[
\frac{{\rm e}^{+\frac{1}{2}-{\cal C}}}{n\pi(a+A)^3}\right]
\label{2Lor3}
\end{eqnarray}
The next Lyapunov exponent should remain at the fluid value (\ref{fluidlyap}):
$\lambda_3\simeq\lambda_{\rm f}<\lambda_2<\lambda_1$.

\begin{figure}
\centerline{\epsfig{file=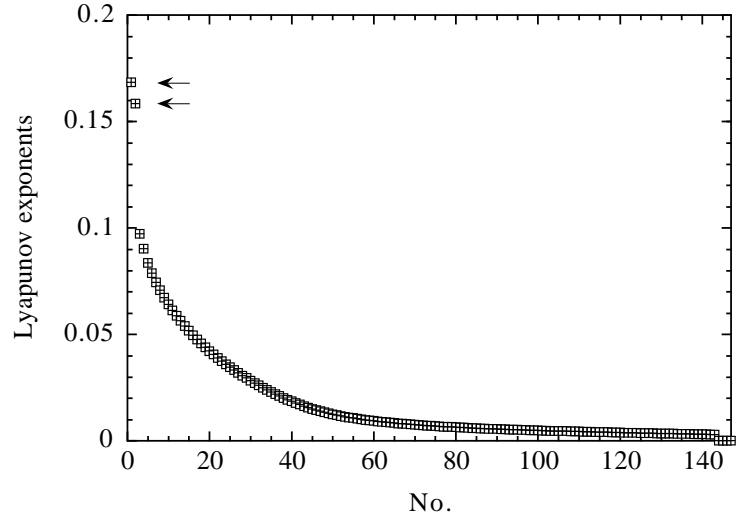,width=10cm}}
\caption{Lyapunov spectrum of a 3d fluid of density $n=10^{-3}$ and temperature
$T=1$ composed of $N_{\rm f}=48$ hard spheres of radius $a=1/2$ and mass
$m=1$ with $N_{\rm
t}=1$ hard sphere of radius $A=1/6$ and mass $M=10^{-2}$.  The squares
depict the positive
exponents and the crosses are minus the negative exponents.}
\label{fig5}
\end{figure}

\begin{figure}
\centerline{\epsfig{file=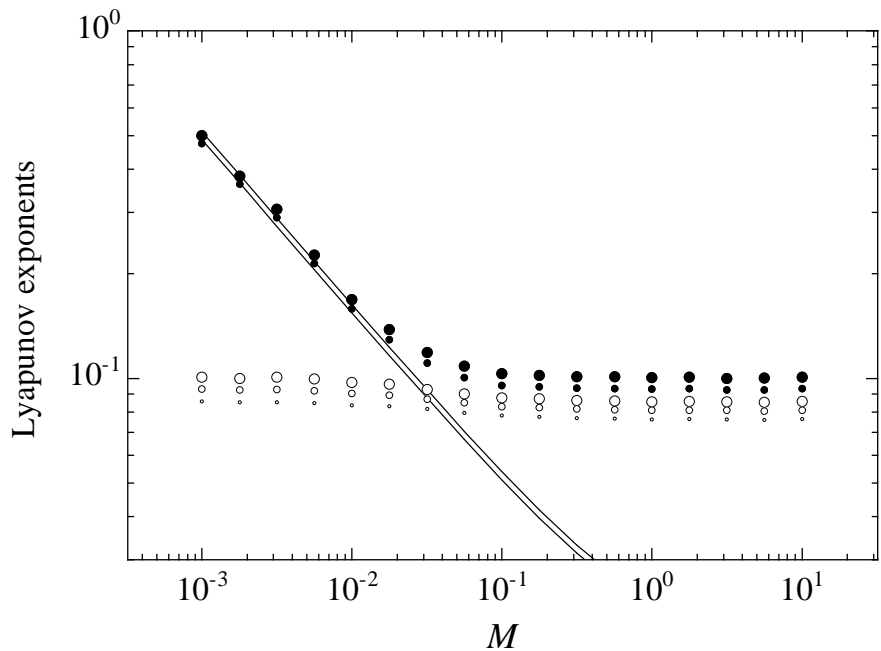,width=12cm}}
\caption{The five largest Lyapunov exponents of a 3d fluid of density
$n=10^{-3}$ and
temperature $T=1$ composed of $N_{\rm f}=48$ hard spheres of radius $a=1/2$
and mass $m=1$ with
$N_{\rm t}=1$ hard sphere of radius $A=1/6$ and varying mass $M$.  The
filled circles depict the
two largest Lyapunov exponents and the open circles the three next ones.
The solid lines are
the predictions of Eqs. (\ref{1Lor3}) and (\ref{2Lor3}).}
\label{fig6}
\end{figure}

Here again, this behavior is
well confirmed by numerical
computation.  Figure
\ref{fig5} depicts a Lyapunov spectrum for $M\ll m$ where we observe that,
indeed, two Lyapunov exponents are separated from the rest of the spectrum,
again with the formation of a gap
as $m/M$ increases.
Figure \ref{fig6} depicts the five
largest Lyapunov exponents as a function of the mass of the tracer particle
showing that
the Lyapunov spectrum is dominated by the tracer particle as soon as $M\ll m$.
The two
largest Lyapunov exponents undergo a
cross-over from the fluid values when
$M \gg m$ to
the Lorentz-gas values (\ref{1Lor3})  and (\ref{2Lor3}) for $M\ll m$.  In
the regime
$M\ll m$, the third and next Lyapunov exponents tend to slightly increase up
to the values of the
first and next Lyapunov exponents of the previous regime $M\gg m$.

\section{The Rayleigh-flight limit}
\label{rayleigh}

Another regime in which we may expect that the tracer particle dominates the
Lyapunov spectrum
is the one
near the limit where the radius of the fluid particles vanishes: $a\to 0$.
This
regime
can be characterized by
the conditions $A\gg a$ and $na^d\ll 1$.  In the limit, the fluid
particles have no collisions
between each other, but only with the tracer particle.  This is
referred to as the
Rayleigh-flight
limit
and is known to present diffusive motion
\cite{rayleigh1891,spohn80}.   In the
limit $a\to 0$
the fluid Lyapunov exponents (\ref{fluidlyap}) vanish.
The tracer particle undergoes elastic, diverging collisions
with the fluid. Since the tracer dynamics
has $d$ degrees of freedom, and the presence of the fluid particles precludes
symmetry transformations of the tracer coordinates alone, we expect
the existence of $d$ positive Lyapunov exponents.  In the
Rayleigh-flight regime, we
may thus expect that the dynamical instability of the fluid is dominated by
the tracer
particle, which is confirmed by the
analysis given below. We want to remark here that the Rayleigh-flight limit
provides an example of a system where the Lyapunov exponents remain
well-defined
in the infinite system limit. In fact taking this limit simplifies the
analysis,
because it largely limits the possibilities of recollisions.

\subsection{Dynamical instability in the Rayleigh flight}
\label{rayleigh.dyn.inst}

Before considering the problem of the full fluid with $a\neq 0$, let us
consider the dynamical
instability of the tracer particle in an infinite domain filled with an
ideal gas of
fluid particles without mutual interaction.  The fluid particles of mass
$m$ are coming from
infinity with velocities distributed according to a Maxwell-Boltzmann
distribution with
temperature $T$ and
have a uniform spatial distribution with density
$n$.  In the limit
$a\to 0$ the fluid particles have an infinite mean free
path.  The only
possible collisions
occur with the tracer particle.  We notice that a fluid particle may
collide more than once with
the tracer particle.  However, recollisions are rare
if either of the conditions $M/m \gg 1$ or $l_{\rm t}/(A+a) \gg 1$ are
satisfied.
It will turn out that in order for the tracer particle to be dominant, at least
one of these conditions has to hold and therefore we will neglect
recollisions in the sequel.

At each collision, the velocity ${\bf v}={\bf v}_{n-1}$ of the fluid
particle is a random vector
of Maxwell-Boltzmann distribution
\begin{equation}
P({\bf v}) = \left(\frac{m}{2\pi k_{\rm B}T}\right)^{\frac{d}{2}} \;
\exp\left(-\frac{m{\bf
v}^2}{2k_{\rm B}T}\right)
\end{equation}
Since the spatial distribution of the fluid particles is uniform the impact
positions are
distributed uniformly over the cross-section.  If $\phi=\phi_n$ denotes the
angle between the
impact unit vector $\hbox{\boldmath{$\epsilon$}}_n$ with the relative
velocity ${\bf
v}_{n-1}-{\bf V}_{n-1}$
\begin{equation}
\cos\phi_n = -\frac{\hbox{\boldmath{$\epsilon$}}_n\cdot({\bf V}_{n-1}-{\bf
v}_{n-1})}{\Vert {\bf
V}_{n-1}-{\bf v}_{n-1}\Vert},
\end{equation}
the uniform distribution of the impact positions implies that
the collision angle
is distributed according to
\begin{eqnarray}
&d=2:& \qquad P(\phi) = \cos\phi
\label{proba.phi.2d} \\
&d=3:& \qquad P(\phi) = \sin 2\phi
\label{proba.phi.3d}
\end{eqnarray}
for $\phi\in\left[0,\frac{\pi}{2}\right]$.
At low density and for $M\gg m$, the successive collisions undergone by the
tracer particle
occur at random time intervals $\tau=\tau_n$ which,
for given speed $V=\Vert{\bf V}_{n-1}\Vert$ of the tracer particle are
distributed
according to
an exponential probability distribution of mean intercollision time $\langle
\tau(V)\rangle=1/\nu_{\rm t}(V)$:
\begin{equation}
P(\tau) = \nu_{\rm t}(V) \; \exp\left[-\nu_{\rm t}(V) \tau\right]
\end{equation}
In the limit $M/m\to\infty$ the velocity dependence of the collision frequency
disappears.

If the recollisions are neglected, the parameters $({\bf v}_{n-1},
\tau_n,\hbox{\boldmath{$\epsilon$}}_n)$ of the successive collisions are
independent random
variables,
though for finite mass ratio their distribution does depend on the tracer
speed $V$.
Therefore, the tracer particle follows a random process which
can be
numerically simulated by a Monte-Carlo method.  The growth
factors of infinitesimal
perturbations on the
position and velocity of the tracer particle also follow a stationary
random process.

Two regimes can be distinguished, depending on the value of the parameter
\begin{equation}
\gamma \equiv \frac{m}{M+m} \; \frac{1}{n(A+a)^d} \sim \sqrt{\frac{m}{M+m}} \;
\frac{\ell_{\rm t}}{A+a}
\label{gamma}
\end{equation}
This parameter is always smaller than the ratio of the mean free path
$\ell_{\rm t}$ of the tracer particle to the collisional radius $A+a$
since $m/(m+M) \le 1$.

\begin{figure}
\centerline{\epsfig{file=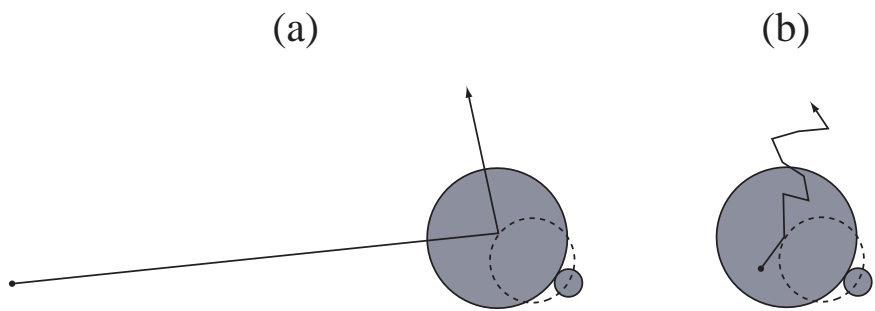,width=12cm}}
\caption{Typical geometry of two successive collisions of the tracer
particle when: (a)
$\ell_{\rm t}\gg A+a$; (b) $\ell_{\rm t}\ll A+a$.  The trajectories are
depicted in the frame
where the fluid particle is at rest.}
\label{fig7}
\end{figure}

For $\gamma \gg 1$ the dynamics in tangent space is very similar to that of a
Lorentz gas.  Typically the position perturbation of the tracer particle when
entering a collision with a fluid particle, can be approximated by the
product of its
velocity perturbation and the free flight time since the previous collision.
Since $\ell_{\rm t}\gg A+a$ in this case,
the dynamical instability is essentially
dominated
by the large distance to the next collision that amplifies the perturbation
on initial
conditions by the ratio $\ell_{\rm t}/(A+a)$ (see
Fig. \ref{fig7}a).  Therefore, the maximal
Lyapunov exponent
can be calculated again by just considering the dynamics from one collision
through the
next one and averaging over the parameters of such an event. In fact,
the Lorentz gas regime may be considered as a special subset of the
class of all systems meeting the requirement $\gamma
\gg 1$. This condition alone, together with the requirement that the
resulting maximal tracer Lyapunov exponent exceeds
the maximal fluid Lyapunov exponent, is sufficient to have dominance of the
tracer particle.

For $\gamma\ll 1$ typically also the relative changes of the perturbations
in a
collision are much smaller than unity.  A possible typical situation is
illustrated in
Fig. \ref{fig7}b in this case.  Therefore, we may describe the dynamics in
tangent space
to a good approximation by a Fokker-Planck equation.  Even though we have
not  found an
analytical solution for this, we can infer several important properties
from it,
especially we can find out how the maximal Lyapunov exponent scales  with
the parameter
$\gamma$.

\vspace*{10mm}
\centerline{\small (a) {\it The regime $\gamma\gg 1$}}
\vspace*{5mm}

In this regime, we can use Krylov's argument \cite{krylov} to estimate the
maximal positive
Lyapunov exponent as follows.  If $\gamma\gg 1$, Eq. (\ref{de}) shows
that a perturbation on the impact unit vector can be estimated as
\begin{equation}
\delta\hbox{\boldmath{$\epsilon$}}_n \sim \frac{\ell_{\rm t}}{A+a}\;
\delta\theta_{n-1}
\end{equation}
where $\delta\theta_{n-1}$ is a perturbation on the angle of the tracer
velocity.  If we
substitute this estimation into Eq. (\ref{dV}) we can conclude that the
perturbation on the
tracer velocity angle should grow
on average as
\begin{equation}
\delta\theta_n \sim \frac{m}{M+m} \, \frac{\ell_{\rm t}}{A+a} \,
\frac{v_{\rm tf}}{v_{\rm t}}
\, \delta\theta_{n-1} \sim \gamma \; \delta\theta_{n-1}
\label{deltatheta}
\end{equation}
in a collision.  Accordingly, the maximal Lyapunov exponent behaves as
$\lambda_{\rm
t}\sim \nu_{\rm t}\ln\gamma$.

The Monte-Carlo simulation confirms this expectation.  Using the last
expression of
Table I in Eq. (\ref{deltatheta}), we find
that the numerical results
for the maximal
Lyapunov exponent of the tracer are well represented by
\begin{equation}
\gamma\gg 1: \qquad \lambda_{\rm t} \simeq \nu_{\rm t} \; \ln
\left[ \frac{m}{M+m}
\; \frac{\alpha_{\rm t}}{n(A+a)^d}\right]
\label{Rlyap1.l>A}
\end{equation}
with the constant $\alpha_{\rm t}=
\exp(1-\frac 3 2{\cal C})/(2\pi^{1/2})=0.322\dots$, in the limit $M\gg m$ for
$d=2$. This agrees well with our simulation results, which give a fitting
value
for $\alpha_{\rm t}$ of $0.33\pm 0.05$.
In the Lorentz-gas limit $M\ll m$, the quantity $\alpha_{\rm t}$
also tends to a constant value, given by $\alpha_{\rm t}={\rm e}^{1-{\cal
C}}/2= 0.763 \dots$ for $d=2$,
as may be seen from Eq. (\ref{Lor2}). The second positive Lyapunov exponent
$\lambda_{\rm t}'$ turns out to have a different dependence on $A+a$
which we have not investigated theoretically yet.

\vspace*{10mm}
\centerline{\small (b) {\it The regime $\gamma\ll 1$}}
\vspace*{5mm}

In this regime, the dynamical instability is weaker.
The perturbation
on average only grows by
a factor slightly larger than unity at each collision. From the Lyapunov
instability,
together with the relationship $\frac{d}{dt}\delta{\bf R}=\delta{\bf V}$,
it is clear that, in the mean, the velocity
and position
perturbations of the Brownian particle are related through
$\delta{\bf V}\sim \lambda_{\rm t} \delta{\bf R}$.
Due to the Brownian fluctuations in the system, $\delta{\bf V}$
will in fact
constantly fluctuate around the value $\lambda_{\rm t}\delta{\bf R}$, both in
magnitude and in direction.
To characterize these fluctuations we
develop a continuous-time description of the dynamical instability similar
to the one
developed by Van Zon in another context \cite{vanzon99}.
We introduce the variables $x$ and ${\bf y}$,
defined
through
\begin{eqnarray}
x&\equiv& \frac{\delta{\bf R}\cdot\delta{\bf V}}{\delta{\bf R}^2}
= \frac{\delta V_{\Vert}}{\Vert\delta{\bf R}\Vert} \label{x.dfn}\\
{\bf y}&\equiv&\frac{ \delta{\bf V}-x\,\delta{\bf R}}{\Vert\delta{\bf R}\Vert}
=\frac{\delta {\bf V}_{\bot}}{\Vert\delta{\bf R}\Vert} \label{y.dfn}
\end{eqnarray}
The variables $x$ and $y\equiv \Vert {\bf y}\Vert$ describe the ratio's of
velocity to position perturbations of
the tracer particle, for the components parallel and perpendicular to
$\delta{\bf R}$
respectively.

They allow us to obtain the maximal Lyapunov exponent as
\begin{eqnarray}
\lambda_{\rm t} &=& \lim_{T\to\infty} \frac{1}{T} \ln
\frac{\Vert\delta{\bf R}(t)\Vert}{\Vert\delta{\bf R}(0)\Vert} \nonumber \\
&=& \lim_{T\to\infty} \frac{1}{T} \int_0^T dt \; \frac{d}{dt}\ln
\Vert\delta{\bf R}(t)\Vert \nonumber \\
&=& \lim_{T\to\infty} \frac{1}{T} \int_0^T dt \; x(t)
\end{eqnarray}
because
\begin{equation}
x = \frac{d}{dt}\ln \Vert\delta{\bf R}\Vert
\end{equation}
and,
again, $\frac{d}{dt}\delta{\bf R}=\delta{\bf V}$.  Accordingly, the maximal
Lyapunov
exponent is given by the time average of the variable $x$.  By ergodicity,
this time average is equal to the statistical average over the stationary
probability
distribution of $x$:
\begin{equation}
\lambda_{\rm t}= \langle x \rangle
\end{equation}

During each free flight between two collisions, the
variables $x$ and $y$ have the
form
\begin{eqnarray}
x &=& \frac{a\tau +b}{a\tau^2+2b\tau+c} \\
y &=& \frac{\pm\sqrt{ac-b^2}}{a\tau^2+2b\tau+c}
\end{eqnarray}
with $\tau-t-t_{n-1}$ and
\begin{equation}
a=(\delta{\bf V}_{n-1})^2 \; , \qquad
b=\delta{\bf R}_{n-1}^{(+)}\cdot\delta{\bf V}_{n-1} \; , \qquad \mbox{and}
\quad
c=(\delta{\bf R}_{n-1}^{(+)})^2
\end{equation}
The ratio $x/y$ is a linear function of time so that
\begin{equation}
K \equiv \frac{d}{dt}\left( \frac{x}{y}\right) = \frac{x^2+y^2}{y}
\end{equation}
is a constant for the time evolution of $x$ and $y$ during the free flights.
Without collisions, the variables $x$ and $y$ thus satisfy equations of motion
\begin{eqnarray}
\dot{x}_{\rm free}&=&-x^2+y^2 = + y^2 \, \partial_yK \nonumber \\
\dot{y}_{\rm free}&=&-2xy = -y^2 \, \partial_xK
\label{xydyn}
\end{eqnarray}

However, the variables $x$ and ${\bf y}$ undergo a jump at each collision
of the
tracer
particle with the fluid particles.  Accordingly, the time evolution of
these quantities
is ruled by the following coupled stochastic equations
\begin{eqnarray}
\dot x &=& -x^2+y^2 + \dot x_{\rm coll} \\
\dot{\bf y} &=& -2x{\bf y} + \dot{\bf y}_{\rm coll}
\end{eqnarray}
with
\begin{eqnarray}
\dot x_{\rm coll} &=& \frac{\delta{\bf R}\cdot\delta\dot{\bf V}_{\rm
coll}}{\delta{\bf R}^2} \label{dx.col}\\
\dot
{\bf y}_{\rm coll} &=&
\frac{\delta\dot{\bf V}_{\rm
coll}-\hbox{\boldmath{$\rho$}}\,(\hbox{\boldmath{$\rho$}}\cdot\delta\dot{\bf V}_
{\rm
coll})}{\Vert\delta{\bf R}\Vert}\, ,
\label{dy.col}
\end{eqnarray}
where we introduced
the unit vector
\begin{eqnarray}
\hbox{\boldmath{$\rho$}} &\equiv& \frac{\delta{\bf R}}{\Vert\delta{\bf
R}\Vert} \label{rho}
\end{eqnarray}
which
is orthogonal
to the unit vector $\hbox{\boldmath{$\sigma$}}\equiv {\bf y}/y$.

The jumps in the perturbation on the velocity are determined by Eq.
(\ref{dV}) in the following form
\begin{equation}
\delta\dot{\bf V}_{\rm coll} = \sum_{n=-\infty}^{+\infty} (\delta{\bf
V}_n-\delta{\bf V}_{n-1}) \; \delta(t-t_n)
\end{equation}
Accordingly, the collisional contributions
(\ref{dx.col})-(\ref{dy.col}) take the forms
\begin{eqnarray}
\dot x_{\rm coll} &=& \sum_{n=-\infty}^{+\infty} \xi_n \; \delta(t-t_n)
\label{dx3.col}\\
\dot
{\bf y}_{\rm coll} &=& \sum_{n=-\infty}^{+\infty}
 \hbox{\boldmath{$ \eta$}}_n \; \delta(t-t_n)
\label{dy3.col}
\end{eqnarray}

To simplify our analysis we assume in the sequel that the mass ratio
$m/(M+m)$ is very small.\footnote{Even if this is not the case our main
results,
notably the way in which the maximal Lyapunov scales with $\gamma$, remain
valid.
But the specific analysis is more complicated, because the simplifications
made
at this point do not apply.} As a result the relative velocity between tracer
and fluid particles may be approximated by minus the velocity of the fluid
particle, the collision frequency of the tracer particle
becomes velocity independent and also the stationary distribution of the
variables $x$ and ${\bf y}$ becomes independent of the velocity of the tracer
particle.
The expressions for the jumps $\xi_n$ and $\hbox{\boldmath{$ \eta$}}_n$
simplify to leading order in
$\frac{m}{M+m}\ll 1$.
In particular, in Eq. (\ref{dV})
the terms
proportional to
$\delta{\bf V}$
may be neglected. In addition, in relative velocities the contributions of
the tracer particle may be ignored. Using Eq.
(\ref{dV}), we obtain,
with these simplifications:
\begin{eqnarray}
\xi_n &\simeq& \frac{2m}{(m+M)(a+A)}\; \Vert{\bf v}_{n-1}\Vert \; \left[
\frac{(\hbox{\boldmath{$\epsilon$}}_n\cdot\hbox{\boldmath{$\rho$}}_n)^2}{\hbox{
\boldmath{$\epsilon$}}_n\cdot{\bf u}_n}
-\hbox{\boldmath{$\epsilon$}}_n\cdot{\bf u}_n\right] \label{xi.approx}\\
\hbox{\boldmath{$ \eta$}}_n &\simeq& \frac{2m}{(m+M)(a+A)}\; \Vert{\bf
v}_{n-1}\Vert \; \left[
\frac{(\hbox{\boldmath{$\epsilon$}}_n\cdot\hbox{\boldmath{$\rho$}}_n)^2}{\hbox{
\boldmath{$\epsilon$}}_n\cdot{\bf u}_n} \, \hbox{\boldmath{$\rho$}}_n
-\frac{\hbox{\boldmath{$\epsilon$}}_n\cdot\hbox{\boldmath{$\rho$}}_n}{\hbox{
\boldmath{$\epsilon$}}_n\cdot{\bf u}_n}
\; \hbox{\boldmath{$\epsilon$}}_n
+(\hbox{\boldmath{$\rho$}}_n\cdot{\bf
u}_n) \, \hbox{\boldmath{$\epsilon$}}_n
-(\hbox{\boldmath{$\epsilon$}}_n\cdot\hbox{\boldmath{$\rho$}}_n) \, {\bf u}_n
\right]\equiv \eta_n \, \hbox{\boldmath{$\tau$}}_n \, ,\label{eta.approx}
\end{eqnarray}
with the unit vector
\begin{equation}
{\bf u}_n \equiv -\frac{{\bf v}_{n-1}}{\Vert{\bf
v}_{n-1}\Vert},
\end{equation}
and with $\hbox{\boldmath{$\tau$}}_n$ a unit vector orthogonal to
$\hbox{\boldmath{$\rho$}}_n$.
Since the successive jumps may be assumed to be statistically independent
random
variables,
the time evolution in tangent space of an ensemble of tracer
particles may
be described by a {\em Boltzmann equation} for the distribution of $x$
and ${\bf y}$, of
the form
\begin{equation}
\partial_tf +\partial_x\left[ (-x^2+y^2) f\right] +\partial_y\left[ (-2xy)
f\right]
=\nu_{\rm t} \int_{-\infty}^{+\infty} d\xi\, \int_0^{\infty} d\eta \,\int'
d \hbox{\boldmath{$\tau$}} \, {\cal
P}(\xi,\eta,\hbox{\boldmath{$\tau$}}) \, \left[ f(x-\xi,{\bf
y}-\eta\hbox{\boldmath{$\tau$}},t)-f(x,{\bf y},t)\right]
\label{Boltz}
\end{equation}
Here ${\cal P}(\xi,\eta,\hbox{\boldmath{$\tau$}})$ is the
distribution function for jumps $\xi$
and
$\eta\hbox{\boldmath{$\tau$}}$ in $x$ respectively ${\bf y}$ as result of a
collision of the tracer particle
with a fluid particle.  The form of this distribution may be obtained from
Eqs.~(\ref{dV})-(\ref{de}), combined
with the Maxwell distribution for the
fluid particle velocity ${\bf
v}_{n-1}$ and
the  distribution of the impact vector
$\delta\hbox{\boldmath{$\epsilon$}}_n$ for given
${\bf V}_{n-1}-{\bf v}_{n-1}$. We used here that the terms in Eq. (\ref{dV})
proportional to
$(\hbox{\boldmath{$\epsilon$}}_n\cdot \delta{\bf V}_{n-1}) \
\hbox{\boldmath{$\epsilon$}}_n$ may be neglected, as result of which
${\cal P}(\xi,\eta,\hbox{\boldmath{$\tau$}})$ becomes independent of $x$
and ${\bf y}$. This is justified
because of our condition $\gamma \ll 1$,
as can be seen on hindsight by comparing this term to the remaining terms in
(\ref{dV}), approximating $\delta{\bf R}$ by $\delta{\bf V}/\lambda_{\rm
t}$ in the
latter [see Eqs. (\ref{xi.approx})-(\ref{eta.approx})].
The prime on the integral over $\hbox{\boldmath{$\tau$}} $ indicates an
integration over the $d-1$-dimensional unit sphere orthogonal to
$\hbox{\boldmath{$\rho$}} $. One readily sees that in fact ${\cal
P}(\xi,\eta,\hbox{\boldmath{$\tau$}})$ as function of
$\hbox{\boldmath{$\tau$}} $ is distributed uniformly over this unit sphere.
Moreover,
this distribution
has long tails due to
so-called {\it grazing collisions}, i.e., collisions with small
scattering angles.  As one sees from the second term in
(\ref{de}) these may give rise to arbitrarily large jumps in $x$ and $y$ of
the order of
\begin{equation}
\xi_n \, , \ \eta_n  \sim \frac{\nu_{\rm t}\, \gamma}{\cos\phi_n}
\end{equation}
As a consequence of the distribution
(\ref{proba.phi.2d}), in $d=2$, or
(\ref{proba.phi.3d}), in $d=3$ the
tails of the probability density
behave as
\begin{equation}
{\cal P}(\xi,\eta,\hbox{\boldmath{$\tau$}} ) \sim \cases{ \frac{\nu_{\rm
t}^2\, \gamma^2}{\vert
\xi\vert^3} \qquad
\mbox{for} \ \vert\xi\vert\to\infty \cr
 \frac{\nu_{\rm t}^2\, \gamma^2}{\vert \eta\vert^3} \qquad \mbox{for} \
\vert\eta\vert\to\infty \cr}
\end{equation}
Therefore the first moments of this distribution exist.  However, the
slow decay of the
distribution for
large jumps leads to logarithmic divergences of the second moments of
${\cal P}$.

When the jumps in $x$ and $\bf y$ at collisions are
mostly small compared to
the typical
magnitudes of $x$ and $\bf y$ themselves, it seems appropriate
approximating the
Boltzmann equation by a Fokker-Planck equation. For this
to be the case we again need the condition
$\gamma \ll 1$.
But in addition, the approximation of the Boltzmann
equation
(\ref{Boltz}) by a Fokker-Planck equation requires the existence of the
second moments
of the  distribution ${\cal P}(\xi,\eta,\hbox{\boldmath{$\tau$}})$.
As we just noticed, these moments do not exist.
But their divergence
is only logarithmic and
therefore the process,
to leading order in $\gamma$, can still be
described by
a Fokker-Planck equation.  The typical scale for the jumps in $x$ and ${\bf y}$
is of order
$\xi,\eta\sim\nu_{\rm t}\gamma$ with $\gamma\ll 1$. Consequently,
if we
introduce a
cut-off in the calculation of the second moment at
values of $\xi$ and $\eta$ satisfying
\begin{equation}
\xi_{\rm cut-off},\; \eta_{\rm cut-off} \; \sim \; \nu_{\rm t} \;
\gamma^{1-\delta},
\label{cut-off}
\end{equation}
with some $\delta >0$,
most of the distribution ${\cal P}$ falls within the cut-offs.
We will justify this cut-off below by an
argument of
self-consistency,
and at the same time determine the appropriate value of $\delta$.
Utilizing the cut-off, we obtain the following Fokker-Planck
equation:
\begin{equation}
\partial_tf +\partial_x\left[ (-x^2+y^2+\mu_{\Vert}) f\right]
+\partial_y\left[ (-2xy) f\right] =D_{\Vert\Vert} \; \partial_x^2 f
+ D_{\bot\bot} \; \partial_{\bf y}^2 f
\label{FP}
\end{equation}
where the drift in $x$ is given by the first moment of the distribution ${\cal
P}$ as
\begin{equation}
\mu_{\Vert} = \nu_{\rm t} \; \langle \xi\rangle\label{mu_x}
\end{equation}
and the ``diffusion coefficients" are defined as
\begin{eqnarray}
D_{\Vert\Vert} &=& \frac{1}{2}\; \nu_{\rm t} \; \langle
(\xi-\langle\xi\rangle)^2\rangle
\label{D_xx}\\  D_{\bot\bot} &=& \frac{1}{2
(d-1)}\; \nu_{\rm t} \; \langle
\hbox{\boldmath{$\eta$}}^2\rangle\label{D_{yy}}
\end{eqnarray}
where $d$ is the space dimension.
There is no drift in $\bf y$ because $\langle\hbox{\boldmath{$\eta$}}\rangle=0$
by symmetry.  For the same reason, the coefficient of
cross-diffusion in $x$ and $\bf y$ vanishes.

Furthermore, in the limit $M/m\to\infty$ we are considering, one has
$\langle\dot{\bf
V}_{\rm coll}\rangle=0$  by isotropy, irrespective of ${\bf V}$ and ${\bf
R}$. Therefore also
$\langle\delta\dot{\bf V}_{\rm coll}\rangle$ and, according to
(\ref{dx.col}) and
(\ref{dx3.col}), $\langle\dot{x}_{\rm coll}\rangle$
and $\langle\xi\rangle$ vanish. A non-zero $\mu_{\Vert}$ may be obtained by
placing the Brownian
particle in a harmonic well, a case we are investigating presently.
>From the expressions (\ref{xi.approx}) and (\ref{eta.approx}) and the cut-offs
(\ref{cut-off}) the diffusion coefficients may be estimated to behave as
\begin{eqnarray}
D_{\Vert\Vert} &=& \frac{1}{2}\; \nu_{\rm t} \; \langle \xi^2\rangle \sim
\nu_t^3 \,
\gamma^2 \, \ln \frac{1}{\gamma}, \label{D_xx.value}\\ D_{\bot\bot} &=&
\frac{1}{2(d-1)}\;
\nu_{\rm t} \;
\langle\hbox{\boldmath{$\eta$}}^2\rangle \sim \nu_t^3 \, \gamma^2 \, \ln
\frac{1}{\gamma},
\label{D_yy.value}
\end{eqnarray}
Note that equation (\ref{FP}) has solutions that are independent of
$\hbox{\boldmath{$\tau$}}$. It is precisely those solutions that we are
interested in.

Continuing with hard spheres we note that, with vanishing first moments, the
Fokker-Planck equation may be rescaled
by
applying the scalings $x\to x\;\kappa^{\frac{1}{3}}$, ${\bf y}\to
{\bf y}\;\kappa^{\frac{1}{3}}$,
and $t \to t\;\kappa^{-\frac{1}{3}}$.
This allows us to extract the
dependence of the
variables on the
expression $\kappa\equiv\nu_{\rm t}^3 \gamma^2
\ln(1/\gamma)$.
The only remaining parameter that
might change on varying the ratio's of masses and diameters is the ratio
$D_{\bot\bot}/D_{\Vert\Vert}$. But even this will change just
very slightly, and in fact only does so because the cut-offs
on $\xi$ and $\eta$ in determining the second moments introduce a
$\gamma$-dependence which works out somewhat differently for
$D_{\Vert\Vert}$ and
$D_{\bot\bot}$.
In the limit $\kappa \to \infty$ this ratio has a well-defined limit.

{F}rom this scaling we may conclude immediately that $\langle x \rangle$, and
hence the
maximal  Lyapunov exponent, scales as $\kappa^{1/3}$ and we finally obtain that
\begin{equation}
\lambda_{\rm t} = \langle x \rangle  \sim \nu_{\rm t} \left( \gamma^2 \;
\ln\frac{1}{\gamma}\right)^{\frac{1}{3}}
\label{Rlyap1.l<A}
\end{equation}
with the parameter
$\gamma$ defined in (\ref{gamma}).

A discussion is now in order to justify the
introduction of the cut-off.  As mentioned already,
the second moments of the distribution
$\cal P(\xi,\eta,\hbox{\boldmath{$\tau$}})$ diverge logarithmically. This
implies that in principle
we should
stick with the Boltzmann equation (\ref{Boltz}). Still, the typical
scale for the jumps in $x$ and ${\bf y}$ is of order $\gamma$, whereas the
typical
scale for variations of $x$ and ${\bf y}$ is expected to be at least
approximately of order $\gamma^{2/3}$, so much larger.
Therefore, the approximation of $f(x-\xi,{\bf y}-\eta
\hbox{\boldmath{$\tau$}} ,t)-f(x,{\bf y},t)$ in
(\ref{Boltz}) by a second order Taylor expansion should still be correct
for values of
$|\xi|$ and $|\eta|$ up to $c\gamma^{2/3}$, with $c$ some small positive
constant.
In addition it can be estimated that the contributions to the Boltzmann
equation from
$|\xi|$- and $|\eta|$-values outside the cut-off are smaller by at least a
factor of
order $(\ln\frac{1}{\gamma})^{1/3}$
than the terms kept in the
Fokker-Planck approximation.
In this way, it is justified to introduce a cut-off of the form
(\ref{cut-off}),
with $\delta=\frac{1}{3}$, which establishes the self-consistency of the
scheme.

\begin{figure}
\centerline{\epsfig{file=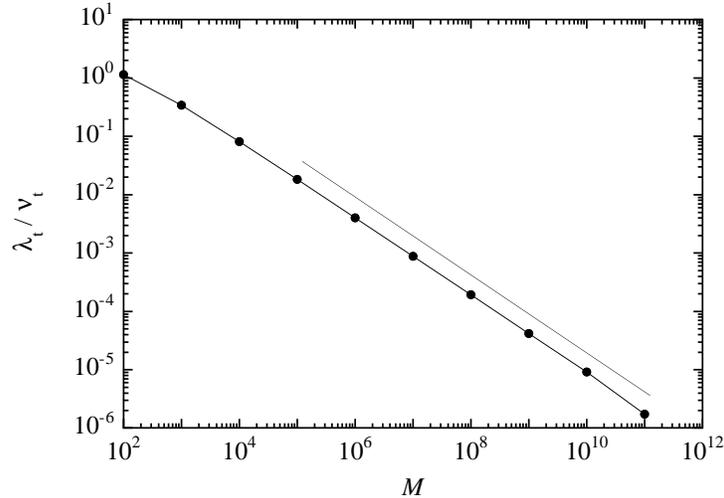,width=10cm}}
\caption{Ratio of the maximal Lyapunov exponent $\lambda_{\rm t}$ to the
collision
frequency $\nu_{\rm t}$ versus the mass $M$ for a tracer disk of radius
$A=499.5$ in
Rayleigh flight in the regime $\gamma \ll 1$.  The Lyapunov exponent is
calculated by Monte-Carlo simulation by assuming that the tracer particle
undergoes independent successive random collisions from fluid disks of radius
$a=0.5$ and mass $m=1$ at density $n=10^{-8}$ and temperature $T=1$.  The
straight line
has the theoretically expected slope $2/3$.}
\label{fig8}
\end{figure}

The numerical computation shows that, indeed, the maximal Lyapunov exponent
is of
the form (\ref{Rlyap1.l<A})
for $\gamma\ll 1$, as seen in Fig.
\ref{fig8}. The logarithmic correction turns out to be very small.  We
further
observed
that the second largest Lyapunov exponent
also scales as
(\ref{Rlyap1.l<A}).

{\bf Remark:} The logarithmic divergences of the second moments of the
distribution $\cal P$ are
an anomaly of the hard-ball potential. For smooth potentials these moments
are well-defined.

\subsection{The full fluid dynamics}

We now consider a fluid of small hard disks of radius $a=\frac{1}{2}$ and
mass $m=1$ with a tracer disk of radius $A$
and mass $M$.  The temperature is always $T=1$.  In order to
find the conditions under which the
tracer particle dominates the Lyapunov spectrum we proceed as follows.

\begin{figure}
\centerline{\epsfig{file=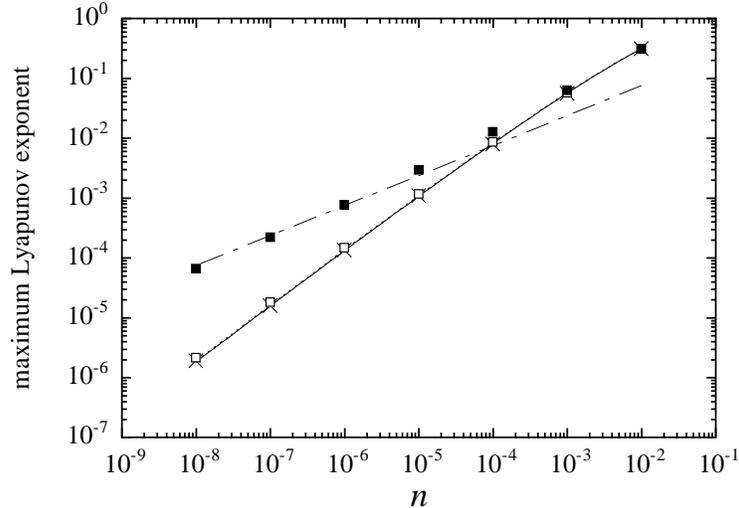,width=10cm}}
\caption{maximal Lyapunov exponent of a 2d fluid of temperature $T=1$ and
varying density $n$
containing $N_{\rm f}\simeq 40$ hard disks of radius $a=1/2$ and mass $m=1$
with one tracer disk
of radius $A$ and mass $M$ in the three cases (\ref{3cases}).  In the
reference case $A=1$ and
$M=1$, the maximal Lyapunov exponent $\lambda_1$ is depicted by the crosses
and the fit by the
solid line.  In the case $A=5$ and $M = 100$, $\lambda_1$ is depicted by
the open squares and
the fit by the dashed line.  In the case $A=1/(2\sqrt{n})$ and $M=10$,
$\lambda_1$ is depicted
by the filled squares.  The long-short dashed line is the fit $\lambda_1 =
0.75\sqrt{n}$.}
\label{fig9}
\end{figure}

We notice that the limit $a\to 0$ is equivalent to the
conditions $A\gg a$ and $n a^d \ll 1$.
Therefore, we consider a sequence of systems of lower and lower density
$n$.  In the
limit $n\to 0$, three types of tracer disks are considered:
\begin{eqnarray}
&(i)& \ A=\frac{1}{2}\, , \qquad M = 1,  \\
&(ii)& \ A=5\, , \qquad M = 100,  \\
&(iii)& \ A=\frac{1}{2\sqrt{n}}\, , \ M = 10.
\label{3cases}
\end{eqnarray}
The specific purpose of the last choice is to consider sequences of
Brownian particles
of increasing sizes, but all made of the same material, so the ratio of $M$
to $nm(a+A)^2$ remains constant.
In Fig. \ref{fig9} the  maximal Lyapunov exponent is plotted versus the
density.

Case (i) is the reference situation in which all
disks are
identical and we recover the
pure fluid behavior (\ref{fluidlyap}),
as expected.  Indeed, a best fit of the
numerical data
to a linear combination of $n\ln n$ and $n$ yields
\begin{equation}
\lambda_{\rm f} = -10.9 \; n \, \ln n - 19.0 \; n,
\label{fluidlyap.nber}
\end{equation}
with the coefficient of the leading term
in agreement with the
value predicted by
Eq. (\ref{fluidlyap}): $-2\sqrt{\pi}\, \omega(41)=-10.89$.

Both cases (ii) and (iii) are
examples of
Brownian motion, as both $A/a$ and $M/m$ are $\gg 1$. In case (ii)
the parameter $\gamma$ changes from $\gg 1$ for the larger of the density
values
considered to $\ll 1$ for the lowest densities. In case (iii) $\gamma$
remains almost fixed at a value of 4/11.
The values of $A/a$ and $M/m$ chosen in case (ii) are such that the fluid
always dominates the maximal Lyapunov exponent.
This is clearly confirmed by the simulation results, which  show that
$\lambda_1 \simeq \lambda_{\rm f}$.

In
case (iii), in contrast, the tracer particle dominates the Lyapunov
spectrum at low
densities as observed in Fig. \ref{fig9}.  Indeed, because
$A=1/(2\sqrt{n})$, we infer from
both Eq.\ (\ref{Rlyap1.l>A})
and (\ref{Rlyap1.l<A}) that
\begin{equation}
\lambda_{\rm t} \sim \nu_{\rm t} \sim \sqrt{n}
\end{equation}
while the fluid Lyapunov exponent takes the values
following from (\ref{fluidlyap.nber}) so
that $\lambda_{\rm f}\ll \lambda_{\rm t}$ for $n\to 0$.
This explains
the behavior observed in Fig. \ref{fig9}.
We notice that
case (iii) is
intermediate between the two
regimes, $\gamma \gg 1$ respectively $\gamma \ll 1$ studied
in subsection \ref{rayleigh.dyn.inst},
because the mean free path of the tracer is
only slightly smaller than
the tracer radius.  Indeed,
in case (iii)
we have $\ell_{\rm t}
\simeq 0.3 \, n^{-\frac{1}{2}} < A+a \simeq 0.5 \, n^{-\frac{1}{2}}$.
Therefore neither (\ref{Rlyap1.l>A}) nor (\ref{Rlyap1.l<A}) strictly applies,
but the scaling of the maximal Lyapunov
exponent as $\sqrt{n}$ remains valid.

However, with a system of one tracer disk among about 40 small disks, we
are still in a
situation very different from the Rayleigh flight because of the periodic
boundary conditions
and the boundedness of the domain.  Even if the mean free path of the fluid
particles among
themselves is much larger than the size of the system $v_{\rm f}/\nu_{\rm
ff} \gg L_x,L_y$
the fluid particles have recollisions on the tracer particle because of the
periodic boundary
conditions.
When $(A/2a)^{d-1} \gg N_{\rm f}$ the dynamics of the fluid particles is
dominated by
recollisions with the tracer particle. We
are then in a situation,
similar to the Sinai billiard
in which a point
particle (such as a fluid particle) collides on a large disk (i.e., the
tracer particle)
in a system with periodic boundaries.  This
leads to two effects.

\begin{figure}
\centerline{\epsfig{file=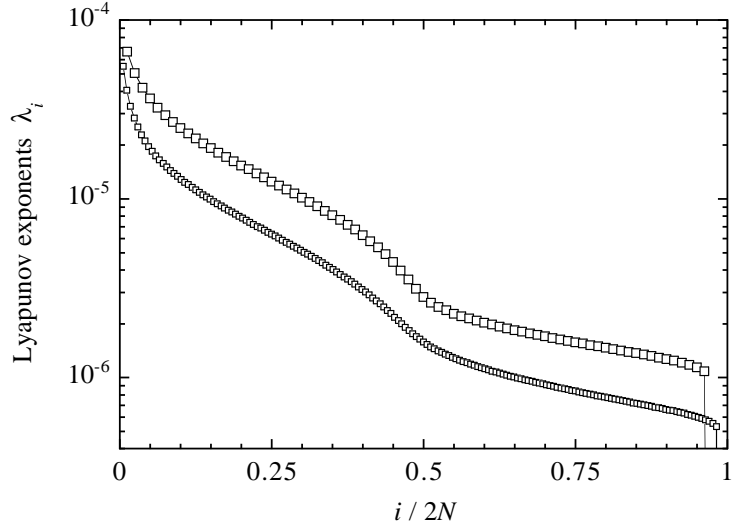,width=10cm}}
\caption{Lyapunov spectra of a 2d fluid of temperature $T=1$ and density
$n=10^{-8}$
containing respectively $N_{\rm f}=39$ (large squares) and $N_{\rm f}=83$
(small squares) hard
disks of radius $a=1/2$ and mass $m=1$ with one tracer disk of radius
$A=5000$ and mass $M=10$
as a function of the relative index $i/(2N)$ of the Lyapunov exponents
$\lambda_i$.}
\label{fig10}
\end{figure}

The first effect is that, in the limit where the tracer particle is very
massive, the fluid
particles become
almost independent of each other.
Therefore the Lyapunov spectrum
becomes similar to
$N_{\rm f}$ copies of the Lyapunov spectrum $(+\lambda_{\rm
S},0,0,-\lambda_{\rm S})$ of the
Sinai billiard.
In the limit $M
\to \infty$, we should thus expect a Lyapunov
spectrum of the
form $(\underbrace{+\lambda_{\rm S},...,+\lambda_{\rm S}}_{N_{\rm
f}},\underbrace{0,...,0}_{2N_{\rm f}},\underbrace{-\lambda_{\rm
S},...,-\lambda_{\rm
S}}_{N_{\rm f}})$ in $d=2$.  This tendency is indeed observed in Fig.
\ref{fig10} which compares
two systems
of different sizes with an identical tracer particle and at the same density.
We
observe in Fig. \ref{fig10} that
indeed the positive Lyapunov exponents
roughly separate
into
two equally
populated families. This tendency is stronger for the 40-particle system
than for the 80-particle one,
as was to be expected in view of the preceding arguments.
In the limit
$L_x,L_y\to\infty$, we notice
that the Lyapunov exponent of the Sinai billiard vanishes as $\lambda_{\rm
S}\sim (v_{\rm
f}/L_{x,y})\ln(L_{x,y}/A)$ so that this effect tends to disappear.

The second effect due to the periodic boundary conditions is more important
for our present considerations
than the previous
one.
If the recollisions of a fluid particle with the tracer particle become
more frequent
than collisions with other fluid particles these may have an important, or
even dominant effect on
the growth of the perturbations
on the tracer
coordinates.
As a
consequence, the maximal
Lyapunov exponent of the system may be different from the Rayleigh-flight value
even though the tracer particle dominates and controls
it.
This
effect
may even remove the
gap in the Lyapunov spectrum, as seen
in Fig. \ref{fig10}.
Notice that a small but finite density of large tracer particles will have
exactly the same type of effects on the
Lyapunov spectrum of a large or even infinite system as periodic boundary
conditions in the case of a single tracer
particle.

\begin{figure}
\centerline{\epsfig{file=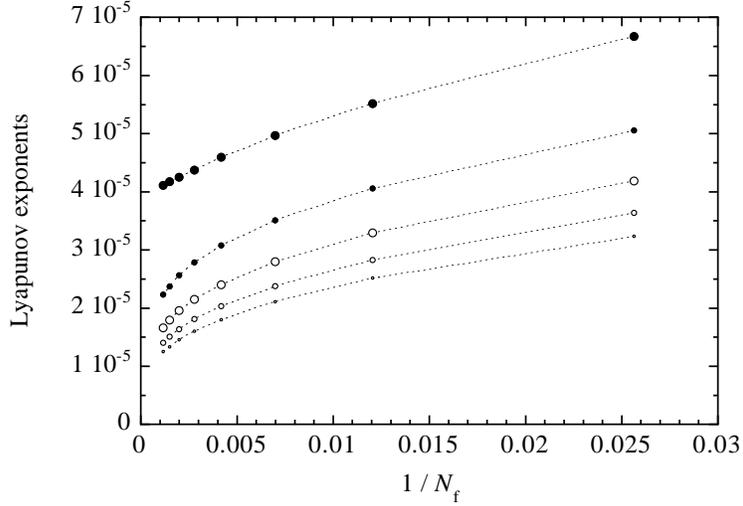,width=10cm}}
\caption{The five largest Lyapunov exponents of a 2d fluid of temperature
$T=1$ and density
$n=10^{-8}$ containing a varying number $N_{\rm
f}=39,83,143,239,359,503,671,863$ of hard disks
of radius $a=1/2$ and mass $m=1$ with one tracer disk of radius $A=5000$
and mass $M=10$ versus
$1/N_{\rm f}$.  The Rayleigh-flight
value of the maximal Lyapunov exponent is
$\lambda_{\rm t}\simeq 3.3 \; 10^{-5}$.}
\label{fig11}
\end{figure}

In the limit where the system becomes infinite, we expect that both effects
disappear and that a
gap
does appear
in the Lyapunov spectrum.  This is indeed the case,
as observed
in Fig. \ref{fig11}
showing the five largest Lyapunov exponents for systems with the same
density $n=10^{-8}$ and
temperature $T=1$, the same tracer particle of radius $A=5000$ and mass
$M=10$, but an
increasing total size and so an increasing number $N_{\rm f}$ of fluid
particles.  Under these
conditions, the limiting Rayleigh flight would have the following two
positive Lyapunov
exponents, obtained by Monte-Carlo simulation:
\begin{eqnarray}
&d=2\, , \ n=10^{-8}\, , \ T=1\, , \ A=5000\, , \ M=10:& \nonumber\\
& \lambda_{\rm t}=(3.3\pm 0.1)
\times 10^{-5} \, , \quad \lambda_{\rm t}'=(0.6 \pm 0.1) \times 10^{-5}&
\end{eqnarray}
For the full dynamics simulated by molecular dynamics, we observe in Fig.
\ref{fig11} that the
maximal Lyapunov exponent decreases as $N_{\rm f}\to\infty$
to a value
close to the maximal
Lyapunov exponent $\lambda_{\rm t}$ of the Rayleigh flight.  The next
exponents decrease
faster,
creating a gap in the Lyapunov spectrum.  We notice that, for the
systems we studied, the
second exponent has not yet converged to the Rayleigh-flight value
$\lambda_{\rm t}'$ and is
therefore not separated from the third and next exponents, but we expect
this to occur for
$N_{\rm f}$ large enough.

In the Rayleigh-flight limit,
$a\to 0$, combined with
the large system limit
$N_{\rm f}\to\infty$, we
therefore
find a Lyapunov spectrum
that is
dominated by the dynamical
instability of the tracer particle:
\begin{equation}
\lambda_1\simeq \lambda_{\rm t} > \lambda_2\simeq \lambda_{\rm t}' \gg
\lambda_3 > \lambda_4 > ...,
\end{equation}
with the formation of a gap,
as in the case of the Lorentz-gas limit,
but
here because of a very
different mechanism.

\section{Conclusions}
\label{conclusions}

In this paper, we have studied a system of hard balls in elastic collisions
(disks in $d=2$ and
spheres in $d=3$) and we have shown that the tracer particle
dominates
the Lyapunov spectrum
in the Lorentz-gas and
in the Rayleigh-flight limit.

In the Lorentz-gas limit, the tracer particle is lighter and moves faster
than the fluid
particles.  The tracer particle
therefore has
a higher collision frequency than
the other particles.
Since the maximal Lyapunov exponent is proportional to the collision
frequency, a gap appears
in the Lyapunov spectrum between the largest Lyapunov exponents,
which are
associated with the
tracer particle,
and the rest of the spectrum. In $d=2$, there is one such
positive Lyapunov
exponent associated with the tracer and, in $d=3$, there are two such
exponents.  These largest
Lyapunov exponents take values very close to the Lorentz-gas values
previously obtained by Van
Beijeren, Dorfman, and Latz \cite{vanbeijeren95,vanbeijeren98}, as
confirmed by direct
numerical computation.

The other limit in which the tracer particle dominates the Lyapunov
spectrum is the
Rayleigh-flight limit.  In this limit, the radius of the fluid particles
tends to zero or,
equivalently, the density of the fluid particles vanishes while the radius
of the tracer is much
larger than the radius of the fluid particles.  In an infinite system, the
only collisions
would occur between the fluid particles
and the lone tracer particle.  We
have shown the
remarkable result that, even in this limit where the fluid is ideal and
composed of
non-interacting particles, the tracer particle may have positive Lyapunov
exponents.
We obtained formulas
for the dependence of the maximal Lyapunov exponent on the parameters of
the system
in two different regimes.
In the first regime, where the mean free path of the tracer particle is so
much larger than  its radius that perturbations
of its coordinates are multiplied by large factors at almost all
collisions, we find a behavior again very similar to that
of the Lorentz gas. In the other regime, where the effect of a single
collision on average is very small, we found that
the maximal Lyapunov exponent scales as the product of the tracer particle
collision frequency and the two-third power of
the mass ratio $m/(m+M)$, up to logarithmic corrections.
These logarithmic corrections are special for the hard-ball potential
and do not occur for more realistic interactions. Moreover, we have shown
that the
positive
Lyapunov exponents determined by the tracer dynamics
may dominate the Lyapunov spectrum of the fully interacting system under
conditions approaching
the Rayleigh-flight limit,
provided the density of tracer particles, or alternatively the ratio
between system size and tracer radius, remains finite.
Again, a gap appears in the Lyapunov spectrum
between the
largest Lyapunov exponents
associated with the tracer particle and
the rest of the
spectrum.

\begin{figure}
\centerline{\epsfig{file=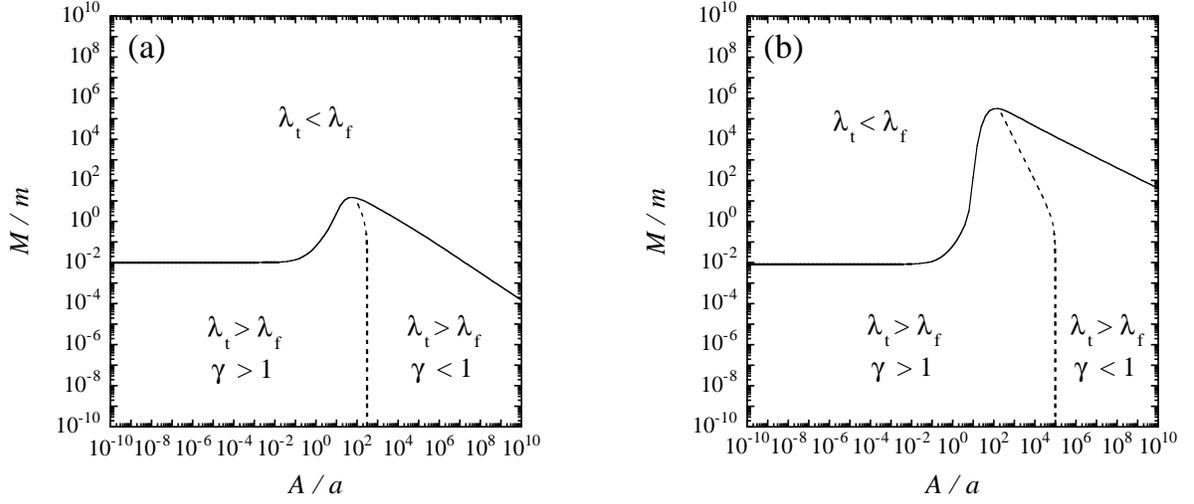,width=16cm}}
\caption{Diagrams of the different regimes of the $d=2$ systems
composed of a tracer particle of radius $A$ and mass $M$
in a fluid of particles of radius $a$ and mass $m$ at density
(a) $n=10^{-5}$ and (b) $n=10^{-10}$ and temperature $T=1$.
The area above the solid line shows the fluid-dominated regime with
$\lambda_{\rm f}>\lambda_{\rm t}$.
The areas below the solid line show the tracer-dominated regimes with
$\lambda_{\rm t}>\lambda_{\rm f}$.
The area on the lower left-hand side is the tracer-dominated regime with
$\gamma >1$,
while the area on the lower right-hand side is the tracer-dominated regime
with $\gamma <1$.
The Lorentz-gas limit is the part of the lower area where $M\ll m$.}
\label{fig12}
\end{figure}

The different tracer-dominated regimes are depicted in Fig. \ref{fig12}
as a function of the mass ratio $M/m$ and the radius ratio $A/a$ for two
different densities
of a two-dimensional fluid. The diagrams are qualitatively similar for a
three-dimensional fluid.
Figure \ref{fig12} shows that the system is in a fluid-dominated regime
if the tracer is very massive.  Nevertheless, we observe in Fig. \ref{fig12}b
that the tracer-dominated regimes extend toward larger masses at lower
densities.

A comment is here in order about Brownian motion.
In typical Brownian motion conditions, the
tracer particle is much more massive than the fluid particles and,
moreover, the mean free path
of the fluid particles among themselves is much shorter than the radius of
the Brownian
particle.  Under such conditions the maximal Lyapunov exponent of the full
system
usually is
essentially the same as the fluid Lyapunov exponent,
as seen in Fig.
\ref{fig12}.
Then the Brownian particle
does not contribute significantly to the dynamical instability of the
system and is a probe for the dynamics of the surrounding fluid.
This is
the case in
typical Brownian-motion experiments.  In order to observe the new effect of
dominance of the dynamical
instability by the Brownian particle
one has to use a
sufficiently rarefied gas as surrounding fluid, in order
to approach to the Rayleigh-flight limit, as seen in Fig. \ref{fig12}b.

\acknowledgments
It is our great pleasure to dedicate this paper to our friend Bob Dorfman
to whom we owe so much.
We want to thank him for sharing with us his interests in the kinetic
theory of Brownian motion,
in the deep connections between chaos theory and statistical mechanics,
in 17th century Dutch painting and in many other important areas of life.
The authors thank Professor G.\ Nicolis for
support and encouragement in this research.
The authors are
supported financially by the National Fund for Scientific Research
(F.\ N.\ R.\ S.\
Belgium), by the Universit\'e Libre de Bruxelles,
and by the
Interuniversity Attraction Pole
program of the Belgian Federal Office of Scientific, Technical and Cultural
Affairs. H.v.B.\ also acknowledges support
by the Mathematical physics program of FOM and NWO/GBE.

\newpage

\end{document}